\documentclass[
twocolumn,
showpacs,
preprintnumbers,amsmath,amssymb,prl]{revtex4-1}

\usepackage{graphicx}
\usepackage{dcolumn}
\usepackage{bm}
\usepackage{color}
\usepackage{multirow}
\usepackage{bbm}
\usepackage{hyperref}
\usepackage{mathptmx}

\begin{document}

\title{Topological insulators, spin, and the tight-binding method}

\author{Warren J. Elder}
\author{Eng Soon Tok}
\altaffiliation{On sabbatical leave from Department of Physics, National University of Singapore, Singapore}
\author{Dimitri D. Vvedensky}
\author{Jing Zhang}
\email{Corresponding email: jing.zhang@imperial.ac.uk}
\affiliation{The Blackett Laboratory, Imperial College London, London SW7 2BW, United Kingdom}


\date{\today}

\begin{abstract}
As one of the first proposed topologically protected states, the quantum spin Hall effect in graphene relies critically on the existence of a spin-dependent gap at the $K/K^\prime$ points of the Brillouin zone. Using a tight-binding formulation based on the method of invariants, we identify the origin of such an intrinsic gap as the {\em three-center interaction} between the $\pi$-orbitals caused by {\em spin-orbit interactions}. This methodology incorporates all symmetry compliant interactions previously neglected and has wider applications for comparisons between first-principle calculations and the tight-binding method. It also identifies a correction to the Haldane model and its generalization,  which incorporates the spin degrees of freedom and reproduces all the salient features required for the quantum spin Hall effect in graphene.
\end{abstract}

\maketitle

The quantum spin Hall effect in graphene is one of the first topologically protected states proposed in any material \cite{kane05}. It depends on two critical elements: the existence of a finite spin-dependent intrinsic gap at $K/K^\prime$ high-symmetry points in the Brillouin zone, and ``band inversion'' and the chiral nature of electronic states in the vicinity of these points. The prediction utilizes a generalization of the Haldane tight-binding model \cite{haldane88} at these points with a spin-dependent term. While the chiral nature of the electronic states is well known from experimental observations of the quantum Hall effect \cite{gusynin05, neto09} and angle-resolved photoemission \cite{liu11} in single-layer graphene, the existence of a {\em finite, spin dependent gap} has not been unequivocally established theoretically. In fact, the absence of a gap without spin, originally proposed by Wallace \cite{wallace47}, was only confirmed by the inclusion of a finite number of higher shells and bases \cite{saito98} using the Slater-Koster \cite{slater54} formulation (SK) of the tight-binding (TB) method. Direct probing by photoemission\cite{ohta07} has not been able to resolve any finite gap. 

The effect of spin-orbit interactions (SOI) on the electronic dispersion of graphene has been investigated using first principles calculation and related to TB with SK formulation with the addition of on-site SOI \cite{min06,yao07} and the incorporation of a $\{d_{xz}, d_{yz}\}$ bases which couples to the $p_z$ states directly \cite{konschuh10}. These studies have shown the existence of a small gap, but did not correctly identify the interaction leading to the spin-dependent gap, nor provide the appropriate generalization of the Haldane model. As will be shown here, the interaction giving rise to the spin-dependent gap is due to three-center interactions neglected by SK.

In this paper the origin of the intrinsic energy gap is examined with the TB method based on the method of invariants \cite{luttinger56}. This implementation of TB includes all symmetry permitted interactions, including three-center terms. Our analysis shows categorically that the symmetry of the crystal ensures both a zero energy gap and linear dispersion at the $K$-point when the SOI is excluded from the Hamiltonian. When the effects of the SOI are included, a small energy gap is predicted to emerge from the inter-site spin-induced mixing of the $\pi$ and $\sigma$ states. The term in the Hamiltonian responsible for this is traced to second-nearest neighbor interactions between $\sigma$ orbitals due to three-center interactions. A small-$k$ expansion at the $K/K^\prime$-point yields a Hamiltonian equivalent to that obtained with the $\bm{k\cdot p}$ method, and shows both the presence of a finite spin-dependent gap, and ``band inversion'' with assoicated chiral nature of the electronic states with an approximately linear dispersion away from $K/K^\prime$. In fact, a simple symmetry-based argument points to the existence of a finite spin-dependent gap. Single-layer graphene has $D_{6h}$ point group symmetry. The dimension of the {\em space group representation} at  $K/K^\prime$ is 4 for $\pi$ bands under the single group, but the largest IR under double group also has dimension of 4. Incorporating spin degrees of freedom must lead to an 8-dimensional reducible space and a gap at these points unless there is an accidental degeneracy, which TB rules out. 
 
In a general TB formulation \cite{slater54}, the Hamiltonian is constructed by summing interactions between L\"owdin orbital states \cite{lowdin50} $|\psi^{\bm{\tau}}_{\xi,p}\rangle$ centered at atomic site $\bm{\tau}$ in the primitive cell and other states $\hat{T}(\bm{R})|\psi^{\bm{\tau}^\prime}_{\eta,q}\rangle$ at equivalent atomic positions $\bm{R}+\bm{\tau}^\prime$ that form a shell of atomic sites around $\bm{\tau}$. These orbitals transform as the irreducible representations (IRs) of the symmetry group of the local bonding configuration (i.e.~containing $s$, $p_x$, $p_y$, and $p_z$ orbitals in graphene) and are constructed so that orbitals centered at different atomic sites are orthogonal. By relocating these orbitals to the origin and denoting them as $|\varphi^{\bm{\tau}}_{\xi,p}\rangle$, the contribution to the Hamiltonian $\hat{H}$ from a given shell is then written as,
\begin{equation}
\label{eqn:shell}
\sum_{\bm{R},\bm{\tau},\bm{\tau}^\prime}^{|\bm{R}+\bm{\tau}^\prime-\bm{\tau}|=R_n}e^{i\bm{k}\cdot (\bm{R}+\bm{\tau}^\prime-\bm{\tau})}\big\langle\varphi^{\bm{\tau}}_{\xi,p}\big|\hat{H}\hat{T}(\bm{R}+\bm{\tau}^\prime-\bm{\tau})\big|\varphi^{\bm{\tau}^\prime}_{\eta,q}\big\rangle\, ,
\end{equation}
where $\bm{R}+\bm{\tau}^\prime-\bm{\tau}$ is directed from the site at $\bm{\tau}$ to that at $ \bm{R}+\bm{\tau}^\prime$ in that shell, and the $\bm{k}$-dependence arises from the exponential functions (EFs).  The contributions to the Hamiltonian from each shell have the full point group symmetry of the crystal. In the SK formulation \cite{slater54} the total Hamiltonian is then constructed by incorporating all interactions from concentric shells of atomic sites in the lattice under the two-center approximation.

While it is perhaps intuitive to work with bases of localized L\"owdin orbitals, they do not always form the basis of IR of the crystal point group. An alternative construction of the Hamiltonian directly utilizes the symmetry of graphene to form bases for the IRs of the crystal point group and the general matrix element theorem \cite{koster58}. One may consider the set of {\em all} equivalent L\"owdin orbital states in the primitive cell relocated to the origin $\{\left|\varphi^{\bm{\tau}}_\eta\right>\}$. They transforms among themselves under the action of the point group of the crystal, and form a generally reducible representation of the group. Using symmetry adapted linear combination (SALC) \cite{cotton90} of $\{\left|\varphi_\eta^\tau\right>\}$, a new set of bases $\{\left|\phi_\mu,i\right>\}$ are constructed which are basis functions for the IR of the crystal point group. The unitary transformation can then be established between the relocated L\"owdin and SALC orbitals.

An invariant form of interactions for a given shell can be constructed for the SALC bases by using symmetrized exponential functions (SEFs) and generators which transform as bases of IRs of the crystal point group. This approach, known as the method of invariants \cite{luttinger56,footnote1}, yields the invariant Hamiltonian from the expression,
\begin{equation}
\label{eqn:invariant_shell}
H_{\mu,\nu}(\bm{k})=\sum_n\sum_\gamma c^{\gamma}_{\mu,\nu}(n)\sum_i {\mathcal{K}^{\gamma}_i(\bm{k},n)}\bigl(M_{\nu,\mu}^{\gamma,i}\bigr)^\dag
\end{equation}
where $n$ indexes the coordinate shells, $\gamma$ indexes the IRs present in those permitted by the general matrix element theorem between states with symmetry $\mu$ and $\nu$, as well as the decomposition of the representation of EFs, $\mathcal{K}^{\gamma}_i(\bm{k},n)$ is the $i$th component from the basis of IR $\gamma$ in the SEF of the $n$th coordinate shell, and $M_{\nu,\mu}^{\gamma,i}$ is the $i$th component of the generator forming the basis of the IR $\gamma $\cite{footnote1}.  The terms $c^{\gamma}_{\mu,\nu}(n)$ represent invariant material parameters which determine the dispersion of a particular crystal with the given symmetry. The existence of a well-defined L\"owdin bases centered at atomic sites places constraints on the invariants. For example, the nearest neighbor interaction ($n=1$), can exist only between appropriate L\"owdin orbital states localized on atomic sites in the graphene crystal. Once the unitary transformation between the L\"owdin and SALC bases is known, the constraints on the invariants can be obtained. The unitary transformation, the form of SEFs for any given shell, and generator matrices can all be obtained using projection operator methods \cite{cotton90,footnote1}.

\begin{figure}[t!]
\includegraphics[width=8.5cm]{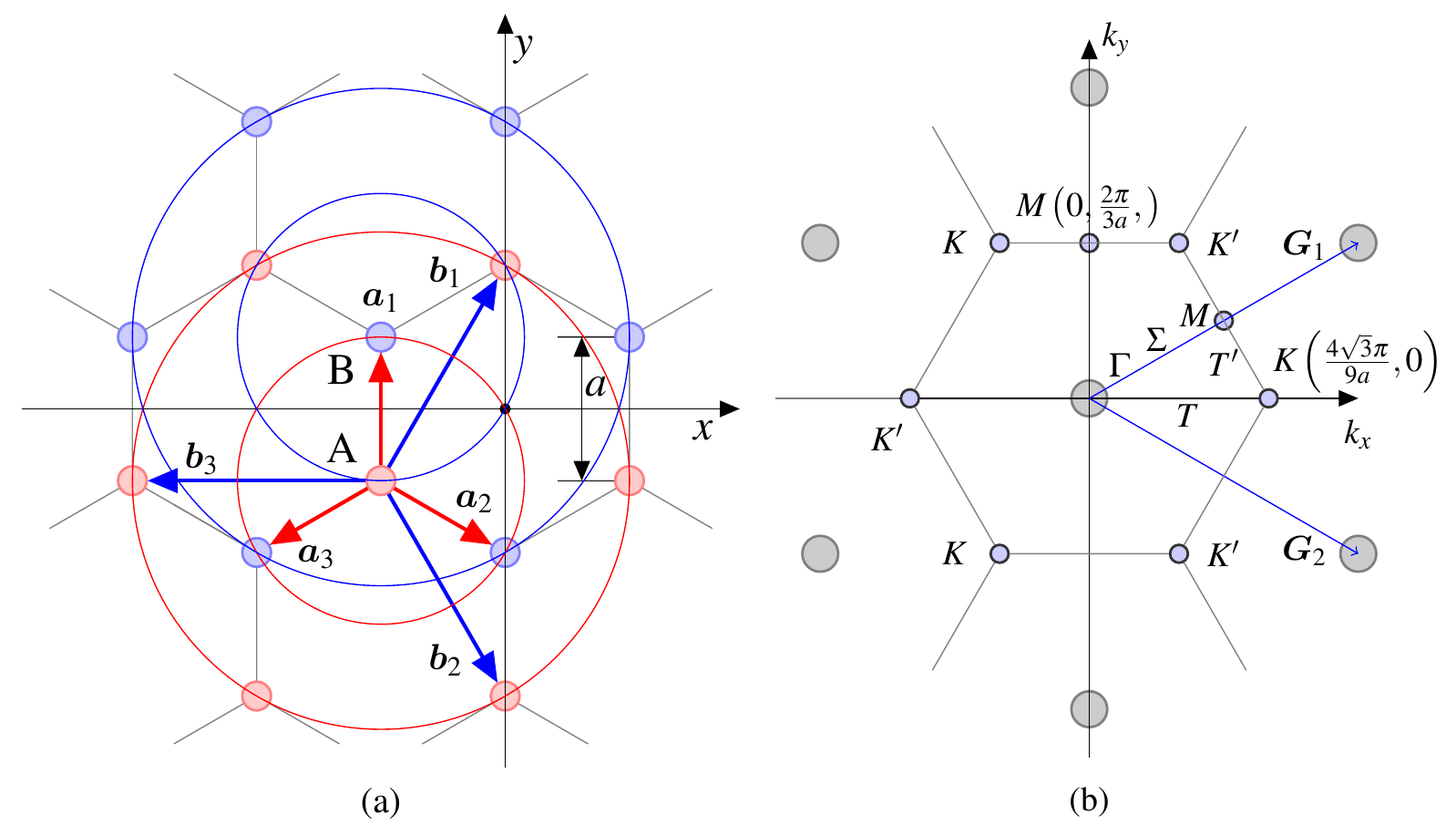}
\caption{(a) Direct space with first and second-nearest neighbor shells in graphene and (b) first Brillouin zone in reciprocal space.}
\label{fig1}
\end{figure}

The real space lattice with nearest neighbor coordinate shells and the first Brillouin zone of graphene are shown in Fig.~\ref{fig1}.  The EFs for a given shell form a closed vector space under the action of the point group and a representation that is generally reducible. The character of a group element in the EF representation is given by the number of EFs left invariant under the action of a group element \cite{dresselhaus08}. The characters of the IRs are readily available from Refs.~\cite{koster63} and \cite{aroyo06}. The symmetry of the allowed SEFs is then obtained from the decomposition of the EF-representation. The on-site interaction generates the trivial $\Gamma_1^+$ IR. For any neighboring shell, the characters of the EF-representation for the $\{E\}$ and $\{\sigma_h\}$ classes are the same and equal to the dimension of the representation. The characters for the $\{C_2^\prime\}/\{\sigma_v\}$ and $\{C_2^{\prime\prime}\}/\{\sigma_d\}$ conjugacy classes are also the same. One of the two pairs can take the value 2 if the axis of the two-fold rotation contains two vectors generating the EFs. The characters of all other classes are zero. The characters for the $\Gamma_3^+, \Gamma_4^+,\Gamma_5^+, \Gamma_1^-, \Gamma_2^-$ and $\Gamma_6^-$  IRs all have opposing signs between pairs of conjugacy classes of $\{E\}/\{\sigma_h\}$, $\{3C_2^\prime\}/\{3\sigma_v\}$ and $\{3C_2^{\prime\prime}\}/\{3\sigma_d\}$. The decomposition theorem then indicates that the SEFs with these symmetries are not permitted in any coordinate shell.

The allowed symmetries of generator matrices which occur between states forming the bases of $\mu$ and $\nu$ can be readily obtained from the decomposition of ${\Gamma^\nu}^\ast\otimes\Gamma^\mu$ \cite{koster58}.  When considering a generator between the same states $(\mu=\nu)$,  time-reversal symmetry must be considered, and only IRs present in the $\left[{\Gamma^\mu}^\ast\otimes\Gamma^\mu\right]_{\rm sym}$ need be considered in the absence of spin. The assignment of symmetry to the energy eigenstates at high symmetry points in the Brillouin zone is made using extended equivalence relations \cite{dresselhaus08} applied to $\pi$ and $\sigma$ orbitals. The symmetry (relating to the co-group of the group of $\bm{k}$) of the Bloch states can then be obtained from the decomposition of such equivalence representations into irreducible components \cite{footnote1,malard09}(see heading in Table~\ref{table1}).

\begingroup
\begin{table}[t!]
\renewcommand{\arraystretch}{1.3}
\caption{Symmetry of permitted generators between SALC basis functions without spin. Those underlined are interactions between $\pi$ and $\sigma$ bonds and have no corresponding SEFs.}
\vskip6pt
\begin{tabular}{|c|cccccc|} \hline
 & $\Gamma_3^+ (\pi)$ & $\Gamma_2^- (\pi)$ & $\Gamma_4^- (\sigma)$ & $\Gamma_5^- (\sigma)$ & $\Gamma_6^+ (\sigma)$ & $\Gamma_1^+ (\sigma)$ \\ \hline
$\Gamma_3^+ (\pi)$ & $\Gamma_1^+$ & $\Gamma_4^-$  & \underline{$\Gamma_2^-$} & \underline{$\Gamma_6^-$} & \underline{$\Gamma_5^+$} & \underline{$\Gamma_3^+$} \\
$\Gamma_2^- (\pi)$ & & $\Gamma_1^+$ & \underline{$\Gamma_3^+$} & \underline{$\Gamma_5^+$} & \underline{$\Gamma_6^-$} & \underline{$\Gamma_2^-$} \\
$\Gamma_4^- (\sigma)$ &  & & $\Gamma_1^+$ & $\Gamma_6^+$ & $\Gamma_5^-$ & $\Gamma_4^-$ \\
$\Gamma_5^- (\sigma)$ & & & & $\Gamma_1^+\oplus\Gamma_6^+$ & $\Gamma_3^-\oplus\Gamma_4^-\oplus\Gamma_5^-$ & $\Gamma_5^-$ \\
$\Gamma_6^+ (\sigma)$ & & & & & $\Gamma_1^+\oplus\Gamma_6^+$   & $\Gamma_6^+$ \\
$\Gamma_1^+ (\sigma)$ & & & & & & $\Gamma_1^+$ \\ \hline
\end{tabular}
\label{table1}
\end{table}
\endgroup

The theory of invariants stipulates that the interaction of a particular symmetry is forbidden unless both the generators and SEFs associated with the same IR exist. Examining the symmetry of generator matrices in Table~\ref{table1}, no interactions can exist between $\pi$ and $\sigma$ bands, since the corresponding product representations decompose into IRs with no corresponding SEFs. We conclude that, without the spin degree of freedom, the interaction between the $\pi$ and $\sigma$ bonding states is forbidden by symmetry.  Examining the SEFs with $\Gamma_4^-$ and $\Gamma_1^+$ symmetry, both clearly return a zero value at the $K$-point for interactions between the A and B sites in relevant neighboring shells. Thus, no interaction occurs between bonding and anti-bonding $\pi$-states at the $K$-point. Moreover, the restriction on the invariants by adherence to localized orbitals requires that the interaction with SEF of $\Gamma_1^+$ symmetry for on-site interaction shells are equal for $\Gamma_3^+$ and $\Gamma_2^-$ states. A zero gap at the $K$-point is then guaranteed in the absence of spin. We have arrived at this conclusion by considering all shells and from symmetry arguments alone, without the need for the two-center approximation in the SK formulation.

The spin-orbit interaction,
\begin{equation}
\hat{H}_{\rm SO}=\frac{\hbar}{4m_0^2c^2}\left(\nabla V(r)\times\hat{\bm{p}}\right)\cdot\hat{\bm{S}}\, ,
\end{equation} 
is a scalar product whose matrix representation with respect to the double group bases transforms according to the trivial representation $\Gamma_1^+$.  This contains both $\bm{k}$-dependent and $\bm{k}$-independent parts and contributes to intra- and inter-site spin-orbit interactions. The symmetry of the permitted generators for interactions between double group bases are obtained using  {\em double group selection rules} and are shown in Table~\ref{table2}. The invariant Hamiltonian can be constructed using Eq.~(\ref{eqn:invariant_shell}) with double group generators and invariants.
\begin{table*}[t!]
\renewcommand{\arraystretch}{1.3}
\caption{Symmetry of permitted generators/operators between all the double group basis functions. Terms with corresponding forbidden SEFs are excluded. Underlined generators have no counterpart in the single group. They are permitted by symmetry only if the spin-orbit interaction is taken into account.}
\begin{center}
\begin{tabular}{|c|cccccccc|} \hline
& $\Gamma_8^+ (\pi)$ & $\Gamma_7^- (\pi)$ & $\Gamma_8^- (\sigma)$ & $\Gamma_9^- (\sigma)$ & $\Gamma_7^- (\sigma)$ & $\Gamma_9^+ (\sigma)$ & $\Gamma_8^+ (\sigma)$ & $\Gamma_7^+ (\sigma)$ \\ \hline

$\Gamma_8^+ (\pi)$ & $\Gamma_1^+$ & $\underline{\Gamma_3^-}\oplus\Gamma_4^-$  & \underline{$\Gamma_5^-$} & \underline{$\Gamma_5^-$} & \underline{$\Gamma_3^-\oplus\Gamma_4^-$} & \underline{$\Gamma_6^+$}  & \underline{$\color{blue}\Gamma_1^+\color{black}\oplus\Gamma_2^+$} & \underline{$\Gamma_6^+$}\\

$\Gamma_7^- (\pi)$ & & $\Gamma_1^+$ &  \underline{$\Gamma_6^+$} &  \underline{$\Gamma_6^+$} & \underline{$\color{blue}\Gamma_1^+\color{black}\oplus\Gamma_2^+$} & \underline{$\Gamma_5^-$} & \underline{$\Gamma_3^-\oplus\Gamma_4^-$} & \underline{$\Gamma_5^-$}\\ 

$\Gamma_8^- (\sigma)$ & & & $\Gamma_1^+$ & $\Gamma_6^+$ & $\Gamma_6^+$ & $\Gamma_5^-$ & $\Gamma_5^-$  & $\Gamma_3^-\oplus\Gamma_4^-$ \\

$\Gamma_9^- (\sigma)$ & & & &$\Gamma_1^+$ & $\Gamma_6^+$ & $\Gamma_3^-\oplus\Gamma_4^-$ & $\Gamma_5^-$ & $\Gamma_5^-$\\

$\Gamma_7^- (\sigma)$ & & & & & $\Gamma_1^+$ & $\Gamma_5^-$ & $\Gamma_3^-\oplus\Gamma_4^-$ & $\Gamma_5^-$\\ 

$\Gamma_9^+ (\sigma)$ & & & & & & $\Gamma_1^+$ & $\Gamma_6^+$ & $\Gamma_6^+$ \\

$\Gamma_8^+ (\sigma)$ & & & & & & & $\Gamma_1^+$ & $\Gamma_6^+$ \\

$\Gamma_7^+ (\sigma)$ & & & & & & & & $\Gamma_1^+$ \\ \hline
\end{tabular}
\end{center}
\label{table2}
\end{table*}

With respect to the double group bases, the $\bm{k}$-independent component of the SOI and the off-diagonal allowed generators with $\Gamma_1^+$ symmetry cause mixing, particularly between the $\pi$ and $\sigma$ bands. For a TB model with a limited basis set, as for a two-band model treating only the bonding and anti-bonding $\pi$ bands, the effects of the remote states (i.e.~$\sigma$ bands) must also be considered. This requires that the states under consideration are decoupled from remote states to the desired order in the SEFs or $\bm{k}$ \cite{lowdin51}. This is similar to how effective mass arises in $\bm{k\cdot p}$ theory with a limited basis. The unitary transformation required to eliminate coupling with remote states introduces additional symmetry-allowed generators. Such generators are not allowed if the double group basis is formed from a direct product of spinor and single-group bases. It is the mixed nature of the bases under consideration that generates additional $\bm{k}$-dependent interactions between the wave functions. In other words, the invariant Hamiltonian formed from the use of double group bases is the most general form of TB Hamiltonian which includes spin-orbit interaction.

\begin{figure}[b!]
\includegraphics[width=8cm]{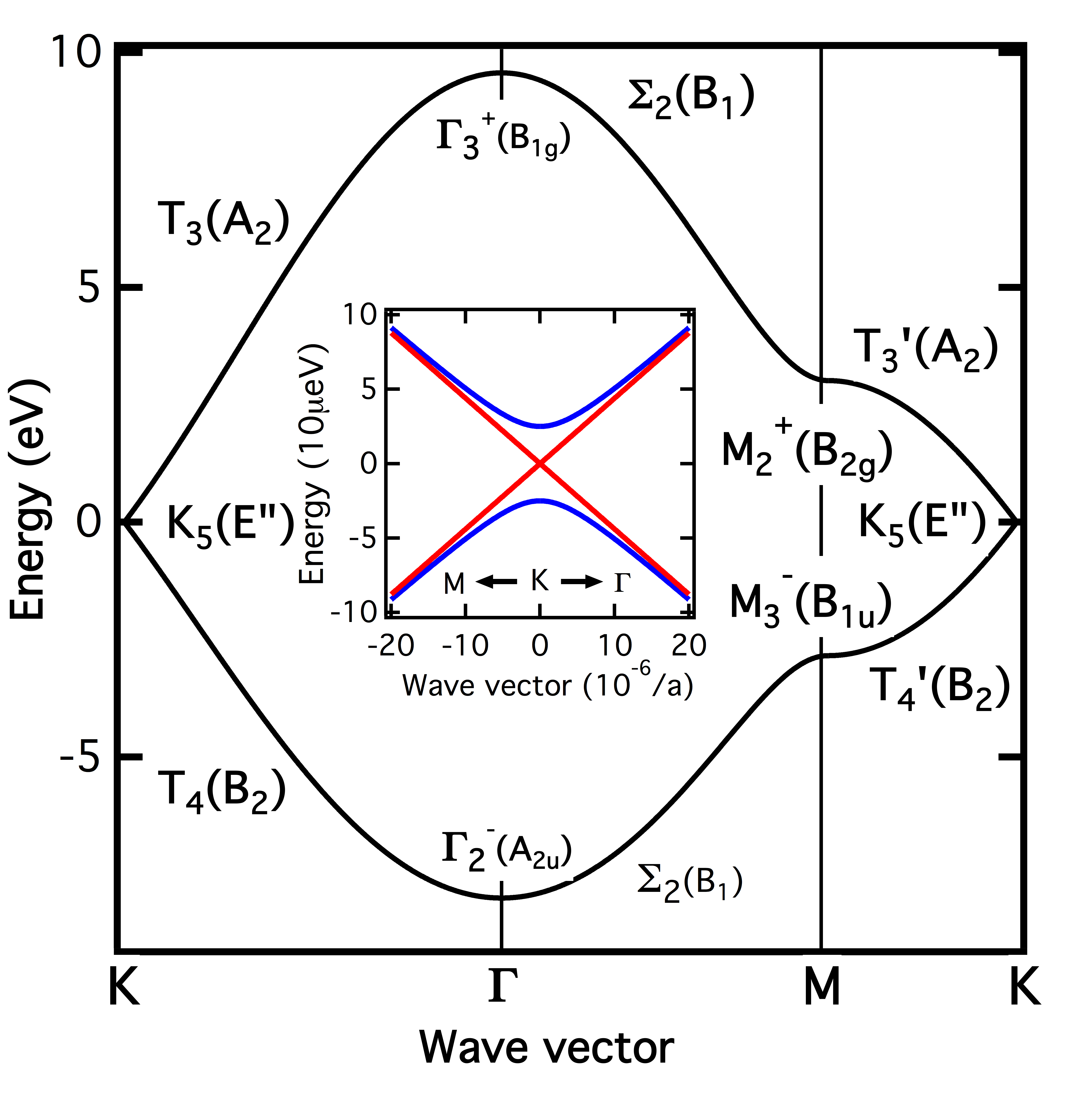}
\caption{Dispersion of graphene $\pi$ band using two-band double-group tight binding method. The insert shows details near the $K$ point with (blue) and without (red) the spin-orbit-interaction $\Gamma_3^-$ term.}
\label{fig2}
\end{figure}

To illustrate this point, we consider a two-band model where only the $\pi$ bands are included directly and the $\sigma$ bands are regarded as remote states. Without electron spin, the only permitted generators between $\pi^\ast$ and $\pi$ are $\Gamma_1^+$ on the diagonal and an off-diagonal $\Gamma_4^-$ (Table~\ref{table1}). To solve the two-band model with spin, terms coupling $\pi$ to $\sigma$ (Table~\ref{table2}) must be eliminated to the desired order in the SEFs. This introduces mixed $\pi$/$\sigma$ bases as a result of SOI. The mixed nature of bases under consideration means interaction between the $\sigma$ orbital states is also reflected in the subspace under consideration, subject to compliance to symmetry. The change in the nearest neighbor parameter does not affect the gap at the $K$-point, since the relevant SEFs are zero at $K$. For second nearest neighbor interaction, the generator of $M_{8+,7-}^{3+}$ is also permitted under double group in addition to those under single group. The corresponding contribution of $\mathcal{K}^{\Gamma_3^-}(\bm{k},2)$ is merely a reflection of the corresponding interaction between $\sigma$ under the SOI induced mixing. The interaction coefficient $c_{8+(\pi),7-(\pi)}^{3-}$ is also purely imaginary in order to preserve the Kramer degeneracy. This leads to the total perturbation $c_{8+(\pi),7-(\pi)}^{3-}\mathcal{K}^{3-}(\bm{k},2)$ being real. The Hamiltonian is constructed using Eq.~(\ref{eqn:invariant_shell}) and the generator matrices 
\begin{equation}
M_{8^+,8^+}^{1^+}=M_{7^+,7^+}^{1^+}=M_{7^-,8^+}^{3^-}=\sigma_0\, ,\quad
M_{7^-,8^+}^{4^-}=\sigma_3\, ,
\end{equation}
where $\sigma_0$ is the $2\times2$ unit matrix and $\sigma_3$ is a Pauli matrix \cite{footnote1}. Figure \ref{fig2} shows the dispersion relations calculated in the two-band model with contributions up to the second-nearest neighbors. The insert shows a gap at the $K$-point arising from the $\mathcal{K}^{\Gamma_3^-}(\bm{k},2)$ term. The relevant SEFs are given in Table~\ref{table3}, together with their Taylor expansions at the $K/K^\prime$-points. 

The size of the gap is determined by the invariant $c_{8^+,7^-}^{3^-}(2)$. From perturbation theory and the mixing scheme, this invariant can be related to the single group interaction between $\sigma$ orbitals associated with $\mathcal{K}^{3^-}(\bm{k},2)$ as 
\begin{equation}
c_{8^+(\pi),7^-(\pi)}^{3^-}(2)=\frac{i\Delta_{\rm SO}^2c_{5^-,6^+}^{3^-}(2)}{(E^{\pi}_{3^+}-E^\sigma_{6^+})(E^\sigma_{5^-}-E^\pi_{2^-})}
\end{equation}
where $\Delta_{\rm SO}^2=c_{8^+(\pi),8^+(\sigma)}^{1+}c_{7^-(\pi),7^-(\sigma)}^{1+}$, are given by the SOI strength. This may be compared with Eq.~(15) of \cite{yao07}. However, $c_{5^-,6^+}^{3^-}(2)$ is a term due to three-center interactions \cite{stiles97}, which is absent in the SK formulation. The corresponding hopping parameter differs from the second-nearest neighbor hopping parameter $c_{8^+(\pi),8^+(\pi)}^{1^+}(2)$ of the two-center interaction. The size of the gap, compared to the  spin-orbit splitting of carbon in diamond \cite{carrier04}, is entirely plausible when considering the origin of the term from SOI-induced mixing and the associated second-nearest neighbor interaction mediated by three-center interactions.

If the Hamiltonian is obtained from a Kronecker product of $\sigma_0$ with the single-group Hamiltonian with additions of an on-site spin-orbit interaction, there are no additional SEFs, and the contribution from the $\Gamma_3^-$ term is absent. If  the constraint of bases centered on atomic sites is retained, there is no gap at the $K$ point. This process is equivalent to the use of bases formed from the direct product of spinor and single group bases, and the absence of a gap implies that such direct product bases are not complete.

\begin{table}[t]
\caption{Relevant SEFs for the first two neighboring shells for the two-band model and their first-order Taylor expansions near the $K/K^\prime$ points. `+' refers to $K$ and `-' refers to $K^\prime$ in expansion. }
\begin{center}
\begin{tabular}{|ccc|}\hline
\multicolumn{3}{|c|}{Shell 1}\\ \hline
IR & Symmetrized exponential functions & $K/K^\prime$ points\\ \hline
$\Gamma_1^+$ & $\frac{2}{\sqrt{6}}\bigl[\cos\left(k_ya\right) + 2\cos\left({1\over2}k_ya\right)\cos\bigl({\sqrt{3}\over2}k_xa\bigr)\bigr]$ & $\mp\frac{\sqrt{6}}{2}\kappa_xa$ \\
$\Gamma_4^-$ &$i\frac{2}{\sqrt{6}}\bigl[\sin\left(k_ya\right) -2\cos\bigl(\frac{\sqrt{3}}{2}k_xa\bigr)\sin\left(\frac{1}{2}k_ya\right)\bigr]$ & $i\frac{\sqrt{6}}{2}\kappa_ya$\\ \hline
\multicolumn{3}{|c|}{Shell 2}\\ \hline
$\Gamma_1^+$ & $\frac{2}{\sqrt{6}}\bigl[\cos\left(\sqrt{3}\,k_xa\right)+2\cos\bigl(\frac{\sqrt{3}}{2}k_xa\bigr)\cos\bigl(\frac{3}{2}k_ya\bigr))\bigr]$ & $-\sqrt{\frac{3}{2}}$ \\
$\Gamma_3^-$ & $i\frac{2}{\sqrt{6}}\bigl[\sin\bigl(\sqrt{3}\,k_xa\bigr) -2\sin\bigl(\frac{\sqrt{3}}{2}k_xa\bigr)\cos\bigl(\frac{3}{2}k_ya\bigr)\bigr]$ &$\mp i\frac{3}{\sqrt{2}}$\\ \hline
\end{tabular}
\end{center}
\label{table3}
\end{table}

The Hamiltonian can be expanded locally around the $K/K^\prime$-point to first order in $\bm{\kappa}=\bm{k}-\bm{k}_{K_0}$. Since the change in basis is second-order in the wave vector, this result is analogous to the $\bm{k\cdot p}$ method using basis functions at the $K$-point with remote states taken into account as a perturbation (effective mass). The effective Hamiltonian, after a suitable transformation \cite{footnote1}, may be written as,
\begin{equation}
\label{eqn:expansion}
H(\bm{\kappa})=\pm\left(
\begin{array}{cc}
c_2\sigma_3 &  c_1a(\kappa_x\mp\mathrm{i}\kappa_y)\sigma_0\\
\noalign{\vskip6pt}
c_1a(\kappa_x\pm\mathrm{i}\kappa_y)\sigma_0 & -c_2\sigma_3
\end{array}
\right)\, ,
\end{equation}
where the choice of $\pm$ determines the point of expansion as $K$ or $K^\prime$. The dispersion relation is then given by 
\begin{equation}
\label{eqn:dispersion}
E(\bm{\kappa})=\pm(c_1^2a^2\kappa^2+c_2^2)^{1/2}\quad c_2\ll c_1.  
\end{equation}

The two essential elements required for the quantum spin Hall effect can be clearly identified from Eqs.~(\ref{eqn:expansion}) and  (\ref{eqn:dispersion}). The dispersion relation in Eq.~\eqref{eqn:dispersion} shows that the $\mathcal{K}^{\Gamma_3^-}(\bm{k},2)$ term is responsible for the creation of a spin-dependent gap through the constant $c_2$. The dependence of signs in Eq.~\eqref{eqn:expansion} on the choice of point of expansion shows the chiral nature of the electronic states with eigenstates at $K/K^\prime$ indicate the inverted nature of the band structure. The fermi velocity $v_F=c_1a/\hbar$ is related to the first nearest neighbor hopping parameter. This Hamiltonian refers to the TB bases at the $K/K^\prime$ points \cite{footnote}.  The energy levels are four-fold degenerate at $K/K^\prime$ points, and separated by a spin-induced gap, while the dispersion retains a linear energy dependence in $|\bm{\kappa}|$ away from these points. 

The results obtained so far allows us to make correction and generalization of the Haldane model \cite{haldane88}. The Haldane model introduces a cell-periodic magnetic field which breaks the time-reversal symmetry, but not the space group symmetry \cite{footnote1}. However, any closed hopping path or its associated Berry phase, is not necessarily invariant. The Berry phase terms $\cos(\phi)$ and $\sin(\phi)$ transform according to $\Gamma_1^+$ and $\Gamma_3^-$ respectively under $D_{6h}$. Hence, $\sin(\phi)\sum_i\sin(\bm{k}\cdot b_i)$ transforms according to $\Gamma_3^-\otimes\Gamma_3^-=\Gamma_1^+$. Thus, the $\bm{k}$ dependent part of the third term in Eq.~(1) of \cite{haldane88} shares the same generator as the first term. The corrected Haldane model under the single group is then
\begin{eqnarray}
H(\bm{k})&=&2t_2\sum_{i=1}^3\left[\cos(\phi)\cos(\bm{k}\cdot \bm{b}_i)-\sin(\phi)\sin(\bm{k}\cdot \bm{b}_i)\right]\sigma_0 \notag\\
&&+t_1\sum_{i=1}^3\left[\cos(\bm{k}\cdot \bm{a}_i)\sigma_1+\sin(\bm{k}\cdot \bm{a}_i)\sigma_2\right].
\end{eqnarray}
This can then be extended to include the spin degree of freedom by utilizing an invariant Hamiltonian and the similarity transform as
\begin{eqnarray}
H(\bm{k}&)&=2t_2\sum_{i=1}^3\left[\cos(\phi)\cos(\bm{k}\cdot \bm{b}_i)-\sin(\phi)\sin(\bm{k}\cdot \bm{b}_i)\right]\sigma_0\otimes\sigma_0\notag \\
&&+2t_2^\prime\sum_{i=1}^3\left[\sin(\phi)\cos(\bm{k}\cdot \bm{b}_i)+\cos(\phi)\sin(\bm{k}\cdot \bm{b}_i)\right]\sigma_{3}\otimes\sigma_{3}\notag \\
&& +t_1\sum_{i=1}^3\left[\cos(\bm{k}\cdot \bm{a}_i)\sigma_1\otimes\sigma_0+\sin(\bm{k}\cdot \bm{a}_i)\sigma_2\otimes\sigma_0\right].
\end{eqnarray}
Note the distinction between the second nearest neighbor hopping parameter which breaks the symmetry between electron and hole states ($t_2$), and that which induces the spin-dependent gap ($t_2^\prime$) and their associated generators.

In summary, we have used the method of invariants in the formulation of the TB method for the  analysis of electronic dispersion in graphene.   The general results shows the $\pi$ band dispersion is indeed gapless in the TB model including an infinite number of shells under the  single group, but {\em a finite spin-dependent gap} arises at $K/K^\prime$ points from SOI-induced mixing between the $\pi$ and $\sigma$ bands. The $\mathcal{K}^{\Gamma_3^-}(\bm{k},2)$ term responsible for the spin-dependent gap is identified as a three-center interaction between $\sigma$ bands manifested as a $\pi$-band interaction through the spin-induced mixing. The SK formulation neglects this interaction and cannot produce such a term. The use of product bases to describe the spin-orbit interactions cannot account for the effect of mixing due to remote states. With spin included, the more general double group selection rules become a necessity.  The local expansion of the Hamiltonian around $K/K^\prime$ shows clearly the existence of a finite spin-dependent intrinsic gap, the chiral nature of the electronic states in the vicinity of Dirac points, and ``band inversion'', necessary for the quantum spin Hall effect. The Haldane model has been corrected and extended to include the spin degree of freedom. The methodology of formulating the TB Hamiltonian using method of invariants takes into account all symmetry compliant interactions, and is more appropriate in any comparisons between first principles calculations and TB models \cite{sutton88}.


\begin{thebibliography}{99}
\bibitem{footnote} Bloch states at $K$ and $K^\prime$ (star of $\bm{k}$) together form the bases of IR of the space group. They are the degenerate energy eigenstates as determined by space group symmetry.
\end{thebibliography}
\end{document}


\title{Supplementary information for: Topological insulators, spin, and the tight-binding method}

\author{Warren J. Elder}
\author{Eng Soon Tok}
\altaffiliation{On sabbatical leave from Department of Physics, National University of Singapore, Singapore}
\author{Dimitri D. Vvedensky}
\author{Jing Zhang}
\email{Corresponding email: jing.zhang@imperial.ac.uk}
\affiliation{The Blackett Laboratory, Imperial College London, London SW7 2BW, United Kingdom}

\date{\today}

\begin{abstract}
This supplement provides the details of the calculations used in the accompanying paper, particularly those related to the group-theoretic methods. The information is given to provide details both on the practical implementation of the method of invariants the tight-binding method, and to avoid confusion regarding different conventions and/or their interpretations.
\end{abstract}

\maketitle

\section{Introduction}

The method of invariants can be used to construct the terms in the tight-binding Hamiltonian between bases associated with the irreducible representations (IRs) $\mu,\nu$ of the point group. This can be written, as given in Eq.~(2) of the main paper,

\begin{equation}
\label{eqn:invariant_shell}
H_{\mu,\nu}(\bm{k})=\sum_n\sum_\gamma\sum_{i=1}^{d_\gamma} c^{\gamma}_{\mu,\nu}(n){\mathcal{K}^{\gamma}_i(\bm{k},n)}\bigl(\bm{M}^{\gamma,q,i}_{\nu,\mu}\bigr)^\dag, \tag{M2}
\end{equation}
where $n$ indexes the coordinate shells, $\gamma$ indexes the IRs present in those permitted by the general matrix element theorem between states with symmetry $\mu$ and $\nu$, as well as the decomposition of the representation of exponential functions (EFs), $\mathcal{K}^{\gamma}_i(\bm{k},n)$ is the $i$th component of occurrence of $\gamma$ IR in the symmetrized exponential functions (SEF) of the $n$th coordinate shell, and $M_{\mu,\nu}^{\gamma,i}$ is the $i$th component of occurrence of IR $\gamma$ of generators. The terms $c^{\gamma}_{\mu,\nu}(n)$ denote the invariant material parameters which actually determine the dispersion of a particular crystal with a given symmetry.

This methodology has the following key features:
\begin{enumerate}
\item Automatically takes into account all symmetry permitted interactions, including the three-center interactions \cite{stiles97};
\item All spatial symmetry and time-reversal symmetry (for intra-band blocks) can be discussed under appropriate selection rules due to use of the full point group of the crystal;
\item For a limited basis set, the effect of remote states is included as a perturbation, if second and higher-order shells are included;
\item Capable of dealing with, and incorporating the effects of spin-orbit interaction, using the appropriate double group bases, and implementing fully, double group selection rules.
\end{enumerate}

The purpose of this supplement is to provide background information on the implementation of this particular methodology, and discusses the following key elements:
\begin{enumerate}
\item Proof of method of invariants applied to tight-binding methods;
\item Basis function $\left|\phi_{\mu,i}\right>$, which form bases of the IRs $\mu$ of the crystal point group and are used to construct Bloch sums;
\item Symmetrized exponential functions $\mathcal{K}^{\gamma}_{i}(\bm{k},n)$;
\item Generator matrices  $M_{\mu,\nu}^{\gamma,i}$;
\item The constraints on invariant material parameters $c^{\gamma}_{\mu,\nu}(n)$ ensuring the Hermiticity of the Hamiltonian and localization of basis functions in the absence of spin degrees of freedom;
\item Generalization of the Haldane model.
\end{enumerate}
The following sections of the supplement deal in turn with each of these items.

\section{Proof of the method of invariants in tight-binding method}

In a general tight-binding formulation \cite{slater54}, the basis functions of the Hamiltonian are constructed from Bloch sums based on localized L\"owdin orbitals $\left|\psi^{\bm{\tau}}_\eta\right>$, which are centered at atomic sites $\bm{\tau}$ and form the basis of IR $\eta$ of the local bonding configuration,
\[
\left|\Psi^{\bm{\tau}}_\eta,\bm{k}\right>=\frac{1}{\sqrt{N}}\sum_{\bm{R}}e^{i\bm{k}\cdot (\bm{R}+\bm{\tau})}\hat{T}(\bm{R})\left|\psi^{\bm{\tau}}_\eta\right>,
\]
where $\hat{T}(\bm{R})$ is the translation operator by a lattice vector $\bm{R}$. An equally valid  localized wave function can be the equivalent bonding orbital states. If we consider such L\"owdin orbitals as centered at the origin, $\left|\varphi^{\bm{\tau}}_\eta\right>$, then
\begin{eqnarray*}
\left|\psi^{\bm{\tau}}_\eta\right>&=&\hat{T}(\bm{\tau})\left|\varphi^{\bm{\tau}}_\eta\right>\, ,\\
\noalign{\vskip3pt}
\left|\Psi^{\bm{\tau}}_\eta,\bm{k}\right>&=&\frac{1}{\sqrt{N}}\sum_{\bm{R}}e^{i\bm{k}\cdot (\bm{R}+\bm{\tau})}\hat{T}(\bm{R}+\bm{\tau})\left|\varphi^{\bm{\tau}}_\eta\right>.
\end{eqnarray*}
A general matrix element of the Hamiltonian with respect to these basis functions is given by
\begin{align*}
&\left<\Psi^{\bm{\tau}}_\xi,\bm{k}|\hat{H}|\Psi^{\bm{\tau}^\prime}_\eta,\bm{k}^\prime\right>=\frac{1}{N}\sum_{\bm{R}^\prime}\sum_{\bm{R}}e^{-i\bm{k}\cdot(\bm{R}^\prime+\bm{\tau})}\left<\varphi^{\bm{\tau}}_\xi|\hat{T}^\dag(\bm{R}^\prime+\bm{\tau})\hat{H}\hat{T}(\bm{R}+\bm{\tau}^\prime)\mid\varphi^{\bm{\tau}^\prime}_\eta\right>e^{i\bm{k}^\prime\cdot(R+\bm{\tau}^\prime)} \\
&\qquad=\frac{1}{N}\sum_{\bm{R}^\prime}\sum_{\bm{R}}\left<\varphi^{\bm{\tau}}_\xi\mid\hat{H}\hat{T}(\bm{R}-\bm{R}^\prime+\bm{\tau}^\prime-\bm{\tau})\mid\varphi^{\bm{\tau}^\prime}_\eta\right>e^{i(\bm{k}^\prime-\bm{k})\cdot\bm{R}}e^{i[\bm{k}\cdot(\bm{R}-\bm{R}^\prime)+\bm{k}^\prime\cdot\bm{\tau}^\prime-\bm{k}\cdot\bm{\tau}]}  \\
&\qquad=\frac{1}{N}\sum_{\bm{R}^{\prime\prime}}\left<\varphi^{\bm{\tau}}_\xi\mid\hat{H}\hat{T}(\bm{R}^{\prime\prime}+\bm{\tau}^\prime-\bm{\tau})\mid\varphi^{\bm{\tau}^\prime}_\eta\right>e^{i[\bm{k}\cdot\bm{R}^{\prime\prime}+\bm{k}^\prime\cdot\bm{\tau}^\prime-\bm{k}\cdot\bm{\tau}]} \sum_{\bm{R}}e^{i(\bm{k}^\prime-\bm{k})\cdot\bm{R}}.
\end{align*}
The last sum yields $N$ if $\bm{k}=\bm{k}^\prime+\bm{K}_m$ ($\bm{K}_m$ is any reciprocal lattice vector), and $0$ otherwise. We obtain, by relabeling $\bm{R}^{\prime\prime}$ as $\bm{R}$,
\[
\left<\Psi^{\bm{\tau}}_\xi,\bm{k}|\hat{H}|\Psi^{\bm{\tau}^\prime}_\eta,\bm{k}^\prime\right>=\sum_{\bm{R}}\left<\varphi^{\bm{\tau}}_\xi\mid\hat{H}\hat{T}(\bm{R}+\bm{\tau}^\prime-\bm{\tau})\mid\varphi^{\bm{\tau}^\prime}_\eta\right>e^{i\bm{k}\cdot(\bm{R}+\bm{\tau}^\prime-\bm{\tau})}\delta_{\bm{k},\bm{k}^\prime+\bm{K}_m}
\]
The Hamiltonian is then specified for a given wave vector $\bm{k}$ in the first Brillouin zone. The matrix element is expected to diminish rapidly with increasing magnitude of the argument of the translation operator. The summation over all primitive cells is then partitioned into sums over $\bm{R}, \bm{\tau}, \bm{\tau}^\prime$ such that $|\bm{R}+\bm{\tau}^\prime-\bm{\tau}|=R_n$ for some radius $R_n$ and then over shells of increasing radius $R_n$. The summation over radii may be truncated because of the diminishing magnitude of the matrix elements with the radius. The contribution to the Hamiltonian matrix element from a given shell of radius $R_n$ is then written as Eq.~(1) of the main paper:
\[
H(\bm{k},n)=\sum_{\bm{R},\bm{\tau},\bm{\tau}^\prime}^{|\bm{R}+\bm{\tau}^\prime-\bm{\tau}|=R_n}e^{i\bm{k}\cdot(\bm{R}+\bm{\tau}^\prime-\bm{\tau})}\bigl<\varphi^{\bm{\tau}}_\xi\mid\hat{H}\hat{T}(\bm{R}+\bm{\tau}^\prime-\bm{\tau})\mid\varphi^{\bm{\tau}^\prime}_\eta\bigr> \tag{M1}
\]
This contribution is invariant under the action of the point group of the crystal\cite{footnote}.

The set of relocated equivalent L\"owdin orbitals $\big|\varphi^{\bm{\tau}}_{\eta}\big\rangle$ within the primitive cell transform among themselves under the action of the point group, thus forming a representation of this group. Using the symmetry-adapted linear combination (SALC) \cite{cotton90}, a basis set $\bigl|\phi_\mu^i\bigr>$ is constructed, which also form basis functions of IR $\mu$ of the point group.  We perform a similarity transform from the localized L\"owdin orbitals $\big|\varphi^{\bm{\tau}}_{\eta,q}\big\rangle$ to the symmetry-adapted linear combination bases $|\phi_{\mu,i}\rangle$, and focus on the block of Hamiltonian indexed by IR $\mu$ and $\nu$:
\begin{equation}
\label{eqn:shell}
H_{\mu,\nu}(\bm{k},n)=\sum_{\bm{R},\bm{\tau},\bm{\tau}^\prime}^{|\bm{R}+\bm{\tau}^\prime-\bm{\tau}|=R_n}
e^{i\bm{k}\cdot (\bm{R}+\bm{\tau}^\prime-\bm{\tau})}
\left<\phi_{\mu}\mid\hat{H}\hat{T}(\bm{R}+\bm{\tau}^\prime-\bm{\tau})\mid\phi_\nu\right>\, .
\end{equation}
We need to establish that the two factors in the expression can be expressed as linear combinations of some basis which transforms according to IR of the point group. 
 
The vectors $\{\bm{R+\tau^\prime-\tau}\}$ in a given shell transform among themselves under the action of point group of the crystal and form a representation of this group. The same holds true for set of exponential functions $\{e^{i\bm{k}\cdot (\bm{R}+\bm{\tau}^\prime-\bm{\tau})}\}$. This representation is generally reducible and decomposed into a set of IRs labelled $\mathcal{A}$. Using the projection operator technique \cite{cotton90}, we can obtain a set of basis functions $\mathcal{K}^{\xi,p}(\bm{k},n),\:\xi\in \mathcal{A}$ such that the action of group element $g$ on such function yields a transformation
\begin{equation}
\label{eqn:transf_ef}
{\mathcal{K}^{\xi,p}_i(\bm{k},n)}^\prime=\sum_{j=1}^{d_\xi}\mathcal{D}(g)_{ji}^\xi\mathcal{K}^{\xi,p}_j(\bm{k},n)\, ,
\end{equation}
and we may express a typical exponential function as
\begin{equation}
\label{eqn:gen_ef}
e^{i\bm{k}\cdot (\bm{R}+\bm{\tau}^\prime-\bm{\tau})}=\sum_{\xi\in \mathcal{A},p}^{\mathcal{A}}a_{\xi,p}^{\bm{R},\bm{\tau},\bm{\tau}^\prime}\sum_{i=1}^{d_\xi}\mathcal{K}^{\xi,p}_i(\bm{k},n)\, ,
\end{equation}
where $p$ is the multiplicity of $\xi$ in the decomposition of the representation. There is a distinction between the set of exponential functions here and the wave vector component in the $\bm{k\cdot p}$ method \cite{luttinger56}. The exponential functions transform with the crystal in the tight-binding method, whereas $\bm{k}$ is an external perturbation and part of the coordinate system.

We may write the matrix element
\begin{eqnarray*}
\left<\phi_\mu\mid\hat{H}\hat{T}(\bm{R}+\bm{\tau}^\prime-\bm{\tau})\mid\phi_\nu\right>&=&\bigl<\phi_\nu\mid\left(\hat{H}\hat{T}(\bm{R}+\bm{\tau}^\prime-\bm{\tau})\right)^\dag\mid\phi_\mu\bigr>^\dag \\
&=&\left<\phi_\nu\mid\hat{H}\hat{T}(-(\bm{R}+\bm{\tau}^\prime-\bm{\tau}))\mid\phi_\mu\right>^\dag\, .
\end{eqnarray*}
The operators $\{\hat{T}(-(\bm{R}+\bm{\tau}^\prime-\bm{\tau}))\}$ have the same vector arguments as the exponential functions for a given shell. Since $\hat{H}$ is invariant under the point group of the crystal, the set of operators $\left\{\hat{H}\hat{T}(-(\bm{R}+\bm{\tau}^\prime-\bm{\tau}))\right\}$ also form a representation of the group, which may be decomposed into the same set of IRs in $\mathcal{A}$. A specific operator may be expressed as
\[
\hat{H}\hat{T}(-(\bm{R}+\bm{\tau}^\prime-\bm{\tau}))=\sum_{\zeta\in \mathcal{A},r}^{\mathcal{A}}w_{\zeta,r}\sum_{k=1}^{p_\zeta}\hat{\mathcal{H}}_{\zeta,r}^k\, ,
\]
where $\hat{\mathcal{H}}_{\zeta,r}$ transforms as IR $\zeta$ of the point group of the crystal, and $r$ is the multiplicity in the decomposition of $\{\hat{H}\hat{T}(-(\bm{R}+\bm{\tau}^\prime-\bm{\tau}))\}$. Let $\mathcal{B}$ be set of IRs for which tensor operators with such symmetry are not forbidden by the general matrix element theorem \cite{koster58} (including time-reversal symmetry where appropriate). Then $\left<\phi_\nu\middle|\hat{\mathcal{H}}_{\zeta,r}\middle|\phi_\mu\right>$ is forbidden unless $\zeta \in \mathcal{B}$. Let $\mathcal{C}=\mathcal{A}\bigcap \mathcal{B}$. Hence,
\begin{eqnarray}
\left<\phi_\nu\mid\hat{H}\hat{T}(-(\bm{R}+\bm{\tau}^\prime-\bm{\tau}))\mid\phi_\mu\right>&=&\sum_{\eta\in \mathcal{C},q}^{\mathcal{C}}b^{\bm{R, \tau, \tau^\prime}}_{\eta,q}\sum_{k=1}^{d_\eta}\bm{M}^{\eta,q,k}_{\nu,\mu}\, ,\notag \\
\left<\phi_\nu\mid\hat{H}\hat{T}(-(\bm{R}+\bm{\tau}^\prime-\bm{\tau}))\mid\phi_\mu\right>^\dag&=&\sum_{\eta\in \mathcal{C},q}^{\mathcal{C}}\bigl(b^{\bm{R, \tau, \tau^\prime}}_{\eta,q}\bigr)^\ast\sum_{k=1}^{d_\eta}\bigl(\bm{M}^{\eta,q,k}_{\nu,\mu}\bigr)^\dag\, ,\label{eqn:gen_matrix}
\end{eqnarray}
where $\eta$ is in the intersection ($\mathcal{C}$) of set of IRs in the decomposition of $\left\{\hat{H}\hat{T}(-(\bm{R}_m+\bm{\tau}^\prime-\bm{\tau}))\right\}$ ($\mathcal{A}$) and the set of IRs permitted by the general matrix element theorem between states of IR $\mu$ and $\nu$($\mathcal{B}$). $q$ is the multiplicity of IR $\eta$ in the decomposition of $\Gamma_\nu^\ast\otimes\Gamma_\mu$ under the general matrix element theorem. The generator matrix $\bm{M}^{\eta,q,k}_{\nu,\mu}$ transforms according to
\begin{eqnarray}
{\bm{M}^{\eta,q,k}_{\nu,\mu}}^\prime&=&\sum_{l=1}^{d_\eta}\mathcal{D}^\eta(g)_{lk}\bm{M}^{\eta,q,l}_{\nu,\mu} \notag\, , \\
\noalign{\vskip3pt}
\bigl({\bm{M}^{\eta,q,k}_{\nu,\mu}}^\prime\bigr)^\dag&=&\sum_{l=1}^{d_\eta}\mathcal{D}^\eta(g)_{lk}^\ast\bigl(\bm{M}^{\eta,q,l}_{\nu,\mu}\bigr)^\dag \, .\label{eqn:transf_matrix}
\end{eqnarray}

Using Eq.~(\ref{eqn:gen_ef},\ref{eqn:gen_matrix}), a typical term in Eq.~\eqref{eqn:shell} can be expressed as
\begin{eqnarray*}
e^{i\bm{k}\cdot (\bm{R}+\bm{\tau}^\prime-\bm{\tau})}&&\left<\phi_\mu\mid\hat{H}\hat{T}(\bm{R}+\bm{\tau}^\prime-\bm{\tau})\mid\phi_\nu\right>=\notag \\
&&\sum_{\xi\in \mathcal{A},p}^{\mathcal{A}}a_{\xi,p}^{\bm{R},\bm{\tau},\bm{\tau}^\prime}\sum_{i=1}^{d_\xi}\mathcal{K}^{\xi,p}_i(\bm{k},n)\sum_{\eta\in \mathcal{C},q}^{\mathcal{C}}\bigl(b^{\bm{R},\bm{\tau},\bm{\tau}^\prime}_{\eta,q}\bigr)^\ast\sum_{k=1}^{d_\eta}\bigl(\bm{M}^{\eta,q,k}_{\nu,\mu}\bigr)^\dag
\end{eqnarray*}
Since $H_{\mu,\nu}(\bm{k},n)$ is invariant under the action of the point group element, we have from Eq.~(\ref{eqn:transf_ef},\ref{eqn:transf_matrix}),
\begin{eqnarray}
H_{\mu,\nu}(\bm{k},n)&=&\frac{1}{|G|}\sum_{g\in G}g\circ H_{\mu,\nu}(\bm{k},n) \notag \\
&=&\sum_{\bm{R},\bm{\tau},\bm{\tau}^\prime}\sum_{\xi\in \mathcal{A},p}^{\mathcal{A}}a_{\xi,p}^{\bm{R},\bm{\tau},\bm{\tau}^\prime}\sum_{\eta\in \mathcal{C},q}^{\mathcal{C}}\bigl(b^{\bm{R},\bm{\tau},\bm{\tau}^\prime}_{\eta,q}\bigr)^\ast\sum_{i=1}^{d_\xi}\sum_{k=1}^{d_\eta}\notag \\
&&\quad\quad\quad\frac{1}{|G|}\sum_{g\in G}\sum_{j=1}^{d_\xi}\mathcal{D}^\xi(g)_{ji}\mathcal{K}^{\xi,p}_j(\bm{k},n)\sum_{l=1}^{d_\eta}\mathcal{D}^\eta(g)_{lk}^\ast\bigl(\bm{M}^{\eta,q,l}_{\nu,\mu}\bigr)^\dag\notag \\
&=&\sum_{\bm{R},\bm{\tau},\bm{\tau}^\prime}\sum_{\xi\in \mathcal{A},p}^{\mathcal{A}}\sum_{\eta\in \mathcal{C},q}^{\mathcal{C}}a_{\xi,p}^{\bm{R},\bm{\tau},\bm{\tau}^\prime}\bigl(b^{\bm{R},\bm{\tau},\bm{\tau}^\prime}_{\eta,q}\bigr)^\ast\sum_{i=1}^{d_\xi}\sum_{k=1}^{d_\eta}\sum_{j=1}^{d_\xi}\sum_{l=1}^{d_\eta}\notag \\
&&\quad\quad\quad\frac{1}{|G|}\underbrace{\sum_{g\in G}\mathcal{D}^\xi(g)_{ji}\mathcal{D}^\eta(g)_{lk}^\ast}_{\mbox{G.O.T}}\mathcal{K}^{\xi,p}_j(\bm{k},n)\bigl(\bm{M}^{\eta,q,l}_{\nu,\mu}\bigr)^\dag \notag \\
&=&\sum_{\bm{R},\bm{\tau},\bm{\tau}^\prime}\sum_{\xi\in \mathcal{A},p}^{\mathcal{A}}\sum_{\eta\in \mathcal{C},q}^{\mathcal{C}}a_{\xi,p}^{\bm{R},\bm{\tau},\bm{\tau}^\prime}\bigl(b^{\bm{R},\bm{\tau},\bm{\tau}^\prime}_{\eta,q}\bigr)^\ast\sum_{i=1}^{d_\xi}\sum_{k=1}^{d_\eta}\sum_{j=1}^{d_\xi}\sum_{l=1}^{d_\eta} \notag \\
&&\quad\quad\quad\delta_{\xi\eta}\delta_{ik}\delta_{jl}\frac{1}{d_\eta}\mathcal{K}^{\xi,p}_j(\bm{k},n)\bigl(\bm{M}^{\eta,q,l}_{\nu,\mu}\bigr)^\dag \notag \\
&=&\sum_{\gamma\in \mathcal{F}}^{\mathcal{F}}\sum_{p}\sum_{q}\sum_{l=1}^{d_\gamma}\mathcal{K}^{\gamma,p}_l(\bm{k},n)\bigl(\bm{M}^{\gamma,q,l}_{\nu,\mu}\bigr)^\dag\biggl[\sum_{\bm{R},\bm{\tau},\bm{\tau}^\prime}a_{\gamma,p}^{\bm{R},\bm{\tau},\bm{\tau}^\prime}\bigl(b^{\bm{R},\bm{\tau},\bm{\tau}^\prime}_{\gamma,q}\bigr)^\ast\biggr]
\end{eqnarray}
where $\gamma$ index IRs in the set $\mathcal{F}=\mathcal{A}\bigcap \mathcal{C}=\mathcal{A}\bigcap \mathcal{B}$. The great orthogonality theorem (G.O.T.) \cite{dresselhaus08} has been used in the derivation above.

Re-labelling the term in the parenthesis as $c_{\mu,\nu}^{\gamma,p,q}(n)$ and sum over all shells, we have the building block of the Hamiltonian given by:
\begin{equation}
\label{eqn:invariant_shell1}
H_{\mu,\nu}(\bm{k})=\sum_n\sum_{\gamma\in \mathcal{F}}^{\mathcal{F}}\sum_p\sum_q\sum_{i=1}^{d_\gamma} c^{\gamma,p,q}_{\mu,\nu}(n){\mathcal{K}^{\gamma,p}_i(\bm{k},n)}\bigl( M_{\nu,\mu}^{\gamma,q,i}\bigr)^\dag
\end{equation}

This shows that the tight-binding interaction considered explicitly as in Eq.~(1) of the main paper can be constructed using the method of invariants. Other contributions, such as the effect of states not considered explicitly, would also be compliant to this method and leads to changes in the invariant material parameters from those considered explicitly. When states of more than one symmetry are considered, there may be additional constraints on the invariants between different blocks, which are discussed in a later section.

When considering graphene and those interactions up to the second-nearest neighbor, multiplicity indices $p$ of the SEFs and $q$ of generators are `1' and the summation over these indices and the indices themselves can be dropped. Then Eq.~(\ref{eqn:invariant_shell1}) reduces to Eq.~(2) of the main paper.

\section{Character table and representation matrices}
\label{sec:group_method}

\begin{figure}[t!]
\includegraphics[width=12.5cm]{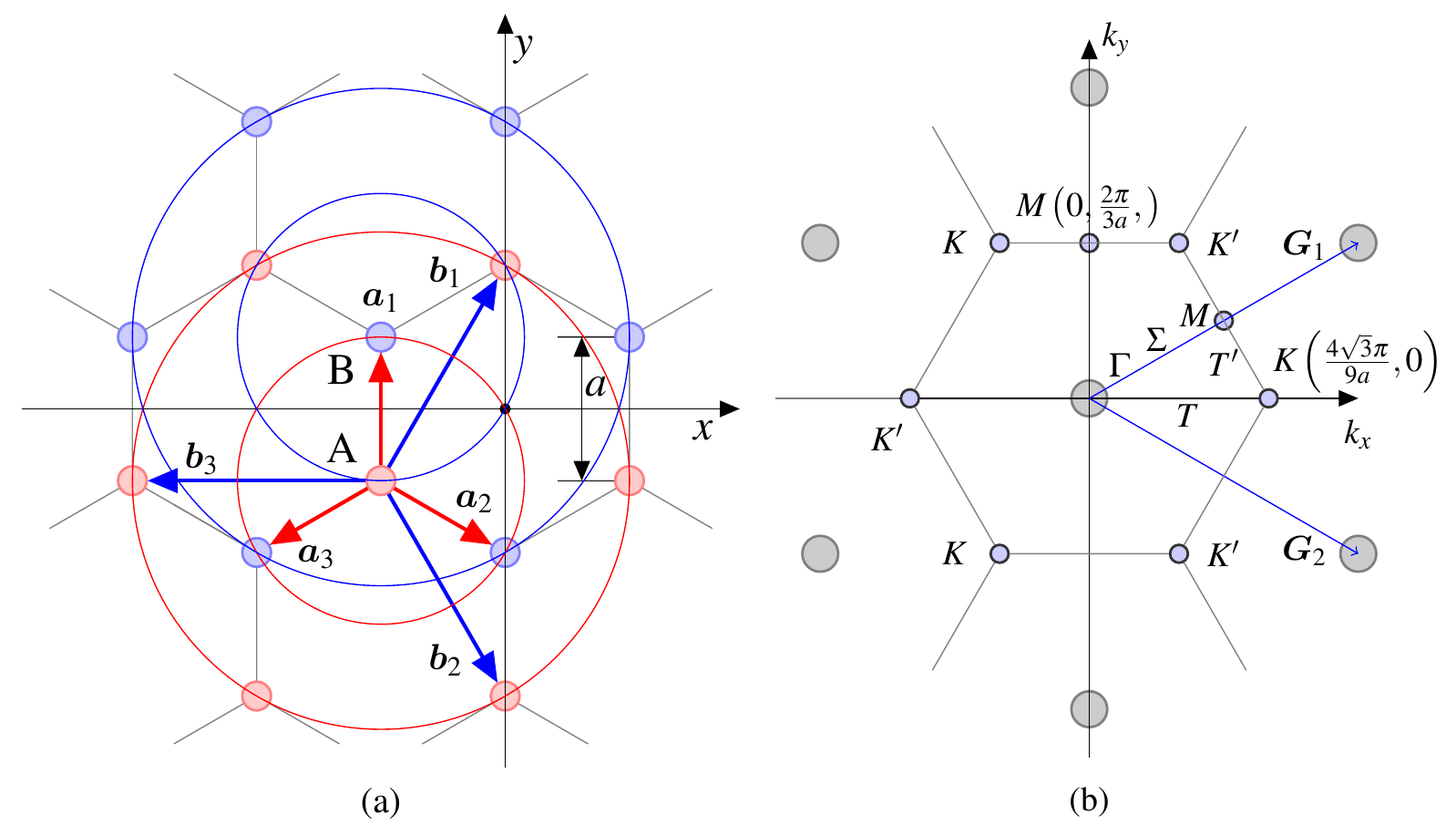}
\caption{(a) Direct space with first and second-nearest neighbor shells of single layer graphene and (b) the first Brillouin zone in reciprocal space.}
\label{fig1}
\end{figure}

The direct and reciprocal space lattices of graphene are shown in Fig.~\ref{fig1}. The factor group of the space group of single layer graphene, with respect to the invariant translation subgroup, is isomorphic to the point group $D_{6h}$. Since the space group is symmorphic, $D_{6h}$ is also a subgroup of the space group. The character table of $D_{6h}$ is given in Table~\ref{tbl:character}, which can be found in \cite{aroyo06,koster63}. The first six single-group conjugacy classes (un-barred operations) are the identity $\{E\}\,(E)$, six-fold rotations about the $z$-axis $\{2C_6\}\,(6)$,  three-fold rotations about the $z$-axis $\{2C_3\}\,(3)$, two-fold rotations about the $z$-axis $\{C_2\} (2_z)$, two-fold rotations about the in-plane axis, including the $x$-axis $\{3C_2^\prime\}\,(2_h)$, and two-fold rotations about the in-plane axis, including $y$-axis $\{3C_2^{\prime\prime}\}\,(2_h^\prime)$.  The remaining six single group conjugacy classes are obtained by the action of the inversion element on the first six classes. The labels of conjugacy classes enclosed in parentheses are those of \cite{aroyo06}. The spinor representation corresponds to the $\Gamma_7^+(E_{3g})$ IR. The use of Mulliken symbols in labelling the IRs follows the convention described in \cite{aroyo06}.
\begin{table*}
\caption{Character table of the point group $D_{6h}=D_6\otimes C_i$.}
\begin{center}
\begin{tabular}{|c|rrrrrrrrr|rrrrrrrrr|}\hline
$D_{6h}$ & $E$ & $\overline{E}$ & $2C_6$ & $2\overline{C_6}$ & $2C_3$ &  $2\overline{C_3}$ & $\begin{array}{c}C_2\\ \overline{C_2}\end{array}$ & $\begin{array}{c}3C_2^\prime\\ 3\overline{C_2^\prime}\end{array}$ & $\begin{array}{c}3C_2^{\prime\prime}\\ 3\overline{C_2^{\prime\prime}}\end{array}$ & $\imath$ & $\overline{\imath}$ & $2S_3$ & $2\overline{S_3}$ & $2S_6$ & $2\overline{S_6}$ & $\begin{array}{c}\sigma_h\\ \overline{\sigma_h}\end{array}$ & $\begin{array}{c}3\sigma_d\\ 3\overline{\sigma_d}\end{array}$ & $\begin{array}{c}3\sigma_v\\3\overline{\sigma_v}\end{array}$\\ \hline \hline
$\Gamma_1^+ (A_{1g})$ &  1 & 1 & 1 & 1 & 1 & 1 & 1 & 1 & 1 & 1 & 1 & 1 & 1 & 1 & 1 & 1 & 1 & 1\\
$\Gamma_2^+ (A_{2g})$ & 1 & 1 & 1 & 1 & 1 & 1 & 1 & $-1$ & $-1$ & 1 & 1 & 1 & 1 & 1 & 1 & 1 & $-1$ & $-1$ \\
$\Gamma_3^+ (B_{1g})$ & 1 & 1 & $-1$ & $-1$ & 1 & 1 & $-1$ & 1 & $-1$ & 1 & 1 & $-1$ & $-1$ & 1 & 1 & $-1$ & 1 & $-1$ \\
$\Gamma_4^+ (B_{2g})$ & 1 & 1 & $-1$ & $-1$ & 1 & 1 & $-1$ & $-1$ & 1 & 1 & 1 & $-1$ & $-1$ & 1 & 1 & $-1$ & $-1$ & 1 \\
$\Gamma_5^+ (E_{1g})$ & 2 & 2 & 1 & 1 & $-1$ & $-1$ & $-2$ & 0 & 0 & 2 & 2 & 1 & 1 & $-1$ & $-1$ & $-2$ & 0 & 0 \\
$\Gamma_6^+ (E_{2g})$ & 2 & 2 & $-1$ & $-1$ & $-1$ & $-1$ & 2 & 0 & 0 &  2 & 2 & $-1$ & $-1$ & $-1$ & $-1$ & 2 & 0 & 0 \\ \hline
$\Gamma_1^- (A_{1u})$ & 1 & 1 & 1 & 1 & 1 & 1 & 1 & 1 & 1 & $-1$ & $-1$ & $-1$ & $-1$ & $-1$ & $-1$ & $-1$ & $-1$ & $-1$ \\
$\Gamma_2^- (A_{2u})$ & 1 & 1 & 1 & 1 & 1 & 1 & 1 & $-1$ & $-1$ &  $-1$ & $-1$ & $-1$ & $-1$ & $-1$ & $-1$ & $-1$ & 1 & 1 \\
$\Gamma_3^- (B_{1u})$  & 1 & 1 & $-1$ & $-1$ & 1 & 1 & $-1$ & 1 & $-1$ &  $-1$ & $-1$ & 1 & 1 & $-1$ & $-1$ & 1 & $-1$ & 1 \\
$\Gamma_4^- (B_{2u})$ & 1 & 1 & $-1$ & $-1$ & 1 & 1 & $-1$ & $-1$ & 1 & $-1$ & $-1$ & 1 & 1 & $-1$ & $-1$ & 1 & 1 & $-1$\\
$\Gamma_5^- (E_{1u})$ & 2 & 2 & 1 & 1 & $-1$ & $-1$ & $-2$ & 0 & 0 &  $-2$ & $-2$ & $-1$ & $-1$ & 1 & 1 & 2 & 0 & 0 \\
$\Gamma_6^- (E_{2u})$  & 2 & 2 & $-1$ & $-1$ & $-1$ & $-1$ & 2 & 0 & 0 & $-2$ & $-2$ & 1 & 1 & 1 & 1 & $-2$ & 0 & 0 \\ \hline
$\Gamma_7^+ (E_{3g})$ & 2 & $-2$ & $\sqrt{3}$  & $-\sqrt{3}$ & 1  & $-1$ & 0 & 0 & 0 & 2 & $-2$ & $\sqrt{3}$  & $-\sqrt{3}$ & 1  & $-1$ & 0 & 0 & 0  \\
$\Gamma_8^+ (E_{4g})$ & 2 & $-2$ & $-\sqrt{3}$ & $\sqrt{3}$  & 1  & $-1$ & 0 & 0 & 0 & 2 & $-2$ & $-\sqrt{3}$ & $\sqrt{3}$  & 1  & $-1$ & 0 & 0 & 0 \\
$\Gamma_9^+ (E_{5g})$ & 2 & $-2$ & 0                & 0               & $-2$ & 2  & 0 & 0 & 0 & 2 & $-2$ & 0                & 0               & $-2$ & 2  & 0 & 0 & 0 \\ \hline
$\Gamma_7^- (E_{3u})$  & 2 & $-2$ & $\sqrt{3}$  & $-\sqrt{3}$ & 1  & $-1$ & 0 & 0 & 0 & $-2$ & 2 & $-\sqrt{3}$ & $\sqrt{3}$ & $-1$  & 1 & 0 & 0 & 0 \\
$\Gamma_8^- (E_{4u})$  & 2 & $-2$ & $-\sqrt{3}$ & $\sqrt{3}$  & 1  & $-1$ & 0 & 0 & 0 & $-2$ & 2 & $\sqrt{3}$  & $-\sqrt{3}$ & $-1$ & 1 & 0 & 0 & 0 \\
$\Gamma_9^- (E_{5u})$  & 2 & $-2$ & 0                & 0               & $-2$ & 2  & 0 & 0 & 0 & $-2$ & 2 & 0                & 0               & 2  & $-2$ & 0 & 0 & 0 \\ \hline
\end{tabular}
\end{center}
\label{tbl:character}
\end{table*}

There are many choices of bases and, correspondingly, many equivalent representation matrices for a given point group. We have used the characters as representation matrices for the one-dimensional IRs. For the $\Gamma_5^-(E_{1u})$ IR, the representation matrices are obtained from the corresponding passive interpretation of the transformation of basis vectors in the $x$- and $y$-directions. Representation matrices of the other, two-dimensional IRs (e.g. $\Gamma_6^+(E_{2u})$), are obtained from the product representation rules $\bigl[\mathcal{D}^{\Gamma_6^+}(g)=\mathcal{D}^{\Gamma_3^-}(g)\mathcal{D}^{\Gamma_5^-}(g)\bigr]$.  The orders of the basis functions in these two-dimensional IRs are determined by the way in which the representation matrices are obtained.  The representation matrices denoted by $\mathcal{D}^\mu(g)$ describe the transformation of basis functions as row vectors, whereas those denoted by $D^\mu(g)$ describe transformation of vector components as column vectors. The passive interpretation is generally used where the action of a group element on a function of space coordinates has the effect,
\begin{equation}
\hat{S}(g)\phi(\bm{r})=\phi(g\bm{r})\, ,
\end{equation}
for $g\in G$.
Representation matrices of the double group IRs are first obtained for $\Gamma_7^+$ (spinor representation) using appropriate generators. Representations of the other IRs of the double group are obtained using appropriate product representations, including block diagonalization when necessary.

\section{Symmetry of equivalent bonding orbitals}

This section establishes the symmetry of basis states formed from equivalent bonding orbitals ($\pi$ and $\sigma$) in graphene, and constructs from them the symmetry adapted linear combination (SALC) bases which form IR of the point group. Under the action of an element of the point group, each set of equivalent bonding orbitals  ($\pi$ and $\sigma$) transform among themselves. They form a closed vector space, and are bases of the representations ($\Gamma^\pi$ or $\Gamma^\sigma$) of the point group (when the point group operation takes an orbital outside the primitive cell, it is translated back into the primitive cell by a lattice vector). These representations are a form of equivalence representation and are generally reducible. In analogy to the equivalence representation of atoms in a primitive cell as defined by Dresselhaus \cite{dresselhaus08}, the character $\Gamma^\beta$ of a set of equivalent bonding orbitals $\{\left|\beta^\tau_r\right>\}$ may be written as,
\begin{equation}
\chi^{\Gamma^\beta}_{\bm{k}}(g)=\sum_{\left|\beta^\tau_r\right>}\delta_{g\left|\beta^\tau_r\right>,\left|\beta^\tau_r\right>}\, ,
\end{equation}
where the summation is over all equivalent orbitals in the primitive cell indexed by $\bm{\tau}$ and $r$, and
\begin{equation}
\delta_{g\left|\beta^\tau_r\right>,\left|\beta^\tau_r\right>}=
\begin{cases}
\hskip9pt 1 & \mbox{~if~}g\left|\beta^\tau_r\right>=\left|\beta^\tau_r\right>\,;\\
\hskip9pt 0 & \mbox{~if~}g\left|\beta^\tau_r\right>\ne\left|\beta^\tau_r\right>\,;\\
-1 & \mbox{~if~}g\left|\beta^\tau_r\right>=-\left|\beta^\tau_r\right>\, .
\end{cases}
\label{eqn:equivalence_rep}
\end{equation}
Here $g$ is an element of the point group. Using this representation, the characters of the representations of $\pi$ and $\sigma$ bonding orbitals are decomposed into IRs of the corresponding point group of the wave vectors. These are shown in Table~\ref{tbl:equiv_orbital}.

\begin{table*}
\caption{Decomposition of equivalence representations of bonding orbitals.}
\label{tbl:equiv_orbital}
\begin{center}
\begin{tabular}{|c|c|c|c|c|c|c|c|r|c|r|c|c|c|} \hline
$\Gamma(D_{6h})$ & $E$ & $2C_6$ & $2C_3$ & $C_2$ & $3C_2^\prime$ &$3C_2^{\prime\prime}$ & $\imath$ & $2S_3$ & $2S_6$ & $\sigma_h$ & $3\sigma_d$ & $3\sigma_v$&Decomposition\\  \hline\hline
$\Gamma(\pi)$ 		& 2 & 0 & 2 & 0 & 0 & $-2$ & 0 & $-2$ & 0 & $-2$ & 2 & 0 & $\Gamma_3^+ (B_{1g}) \oplus \Gamma_2^- (A_{2u})$ \\  
$\Gamma(\sigma)$ 	& 6 & 0 & 0 & 0 & 0 &\hskip8pt  2 & 0 &  0 & 0 & 6 & 2 & 0 & $\Gamma_4^- (B_{2u})\oplus \Gamma_5^- (E_{1u}) \oplus \Gamma_6^+ (E_{2g})\oplus \Gamma_1^+ (A_{1g})$ \\  \hline \hline
\end{tabular}
\end{center}
\end{table*}

Since all the equivalent bonding states in the primitive cell form a representation $\Gamma^\beta$ of the point group of the crystal, a set of symmetry-adapted linear combination (SALC) \cite{cotton90} bases $\left|\phi_{\mu,i}\right>$, can be constructed from $\left|\beta^{\bm{\tau}}_{r}\right>$, and form bases of IRs of the point group of the lattice. The unitary transformation between $\left|\beta^{\bm{\tau}}_{r}\right>$ and $\left|\phi_{\mu,i}\right>$ may be obtained using projection operators \cite{cotton90}. 

\begin{figure}[b!]
\includegraphics[width=12.5cm]{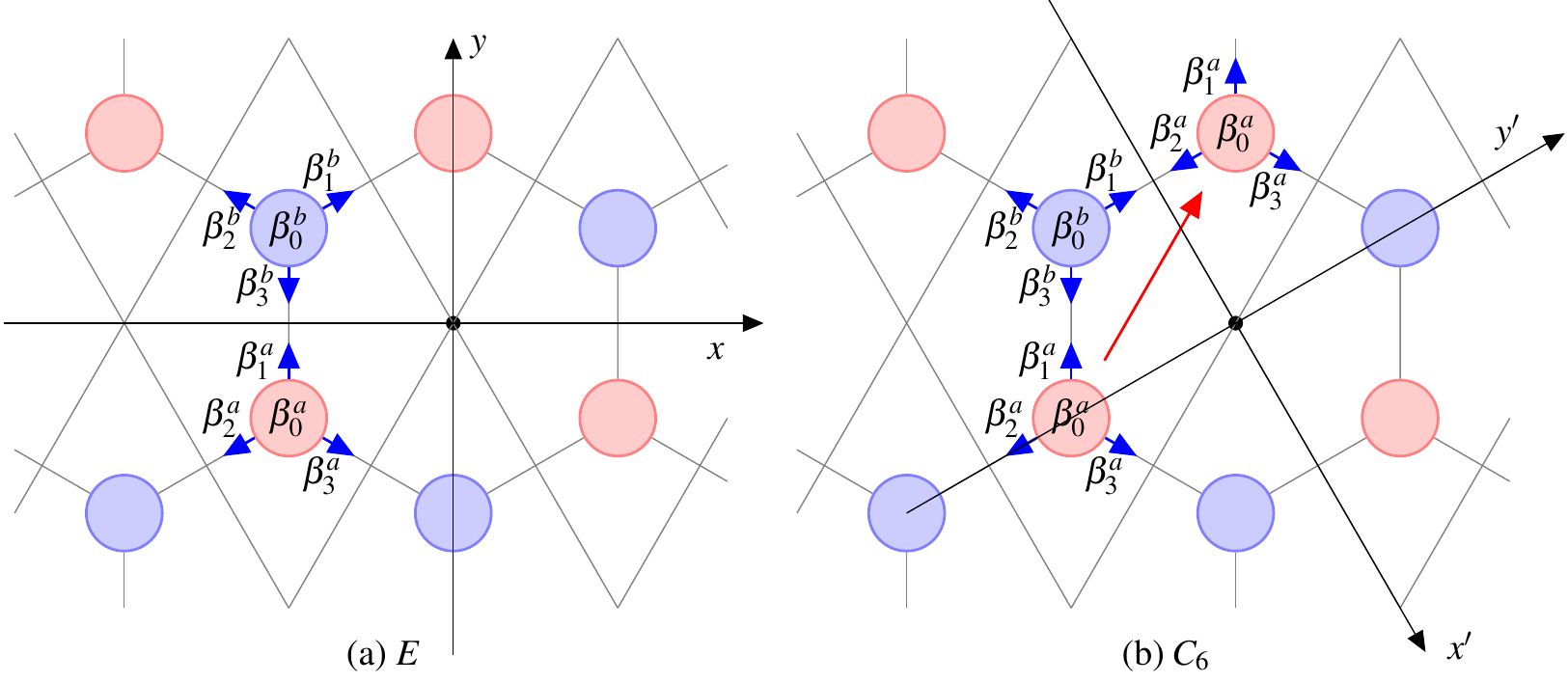}
\caption{(a) Equivalent bonding states in the primitive cell of graphene, and (b) transformation induced by $C_6$ operation (in the passive interpretation).}
\label{fig2}
\end{figure}

The origin in Fig.~\ref{fig1}(a) is the fixed point for operations of the point group. The equivalent bonding states on each of the atomic sites are $\left|\beta^{\{a,b\}}_{\{0\cdots 3\}}\right>$, as shown in Fig.~\ref{fig2}. The superscript $\{a,b\}$ refers to atomic sites A and B, and the subscript $\{0,1,2,3\}$ refers to the $\pi (0)$ and $\sigma (1,2,3)$ bonding states. Under the action of elements of the point group, the $\pi$ bonding states $\left\{\left|\beta^{\{a,b\}}_0\right>\right\}$ and the $\sigma$ bonding states $\left\{\left|\beta^{\{a,b\}}_{\{1\cdots 3\}}\right>\right\}$ form two closed sets which are bases of the representations $\Gamma^\pi$ and $\Gamma^\sigma$ of $D_{6h}$.
Figure~\ref{fig2}(b) shows the transformation induced by the $C_6$ operation using passive interpretation, under which the transformation of bases, described by the representation matrix $\mathcal{D}^\beta(C_6)$, is given by
\begin{subequations}
\begin{equation}
\left(\left|\vphantom{\beta^b}\beta^{a}_{0}\right>^\prime,\left|\beta^{b}_{0}\right>^\prime\right)=\left(\left|\vphantom{\beta^b}\beta^{a}_{0}\right>,\left|\beta^{b}_{0}\right>\right)\begin{pmatrix} 0 &\hskip6pt 1\\ 1 &\hskip6pt 0 \end{pmatrix}
\end{equation}
\begin{eqnarray}
&\left(\left|\beta^{a}_{1}\right>^\prime,\left|\beta^{a}_{2}\right>^\prime,\cdots, \left|\beta^{b}_{3}\right>^\prime\right) =\left(\left|\beta^{a}_{1}\right>,\left|\beta^{a}_{2}\right>,\cdots, \left|\beta^{b}_{3}\right>\right) 
\begin{pmatrix} 0 &\hskip6pt 0 &\hskip6pt 0 &\hskip6pt 0 &\hskip6pt 1 &\hskip6pt 0 \\ 0 &\hskip6pt 0 &\hskip6pt 0 &\hskip6pt 0 &\hskip6pt 0 &\hskip6pt 1 \\ 0 &\hskip6pt 0 &\hskip6pt 0 &\hskip6pt 1 &\hskip6pt 0 &\hskip6pt 0 \\ 1 &\hskip6pt 0 &\hskip6pt 0 &\hskip6pt 0 &\hskip6pt 0 &\hskip6pt 0 \\ 0 &\hskip6pt 1 &\hskip6pt 0 &\hskip6pt 0 &\hskip6pt 0 &\hskip6pt 0 \\ 0 &\hskip6pt 0 &\hskip6pt 1 &\hskip6pt 0 &\hskip6pt 0 &\hskip6pt 0 \end{pmatrix}
\end{eqnarray}
\end{subequations} 

If we relocate all these equivalent orbital states to be centered at origin, they will transform among themselves in the same way as given by the representation matrices above. A general state centered at origin is expressed in terms of these bases as,
\begin{equation}
\left|\Psi\right>=\sum_{\tau, r}c^\tau_r\left|\beta^\tau_r\right>\, ,
\end{equation}
where the summation is over all equivlent obital states in the primitive cell indexed by $\bm{\tau}$ and $r$. Then, the components $c^\tau_r$, viewed as a column vector, transforms according to,
\begin{equation}
c^{\tau^\prime}_{r^\prime}=\sum_{\tau, r}D^{\beta}(g)_{\tau^\prime,r^\prime;\tau,r}c^{\tau}_{r}\, ,
\end{equation}
where $D^\beta(C_6)={\mathcal{D}^\beta(C_6)}^\dag$. Representation matrices of all elements of $D_{6h}$ are constructed in the same way. The projection operator can be defined in terms of these matrices as,
\begin{equation}
\mathcal{P}^\mu_{ij}=\frac{d_\mu}{|G|}\sum_g^{D_{6h}}\mathcal{D}^\mu(g)_{ij}^\ast D^\beta(g)\, ,
\label{eqn:projection}
\end{equation}
where $\mathcal{D}^\mu(g)$ is the matrix of IR $\mu$, and $|G|=24$ is the order of $D_{6h}$. The SALC bases can then be constructed using this projection operator, and then expressed in terms of the linear combination of equivalent bonding orbital states $\{\left|\beta^\tau_r\right>\}$ centered at origin. A similar procedure can be carried out to relate the L\"owdin orbital states $|\varphi^{\bm{\tau}}_{\xi,p}\rangle$, to the equivalent bonding states $\{\left|\beta^\tau_r\right>\}$, using the symmetry group of the bonding configuration of the atomic site. Therefore, there is a unitary transformation between the SALC $\left\{\left|\phi_{\mu,i}\right>\right\}$ centered at origin and L\"owdin basis $\left\{\left|\varphi^{\bm{\tau}}_{\xi,r}\right>\right\}$ centered at origin,
\begin{equation}
\label{eqn:lowdin2salc}
\left|\phi_{\mu,i}\right>=\sum_{\tau,\xi,r}U_{\tau,\xi,r; \mu,i}\left|\varphi^{\bm{\tau}}_{\xi,r}\right>
\end{equation}
For the $\pi$ bands, one obtains
\begin{subequations}
\begin{equation}
\left(\left|\phi_{\Gamma_3^+}\right>,\left|\phi_{\Gamma_2^-}\right>\right)=\frac{1}{\sqrt{2}}\left(\left|p_z^A\right>,\left|\vphantom{p_z^A}p_z^B\right>\right)\begin{pmatrix}\hskip12pt 1  &\hskip12pt  1  \\ \hskip2pt-1 &\hskip12pt 1 \end{pmatrix}
\end{equation}
For the $\sigma$ bands, one obtains
\begin{eqnarray}
&&\left(\left|\phi_{\Gamma_4^-}\right>,\left|\phi_{\Gamma_5^-,1}\right>,\left|\phi_{\Gamma_5^-,2}\right>,\left|\phi_{\Gamma_6^+,1}\right>,\left|\phi_{\Gamma_6^+,2}\right>,\left|\phi_{\Gamma_1^+}\right>\right)=\notag \\
&&\quad\quad\frac{1}{\sqrt{2}}\left(\left|s^A\right>,\left|p_x^A\right>,\left|p_y^A\right>,\left|\vphantom{s^A}s^B\right>,\left|\vphantom{s^A}p_x^B\right>,\left|\vphantom{s^A}p_y^B\right>\right)\begin{pmatrix}
\hskip12pt  1  &\hskip12pt  0  &\hskip12pt  0  &\hskip12pt  0  &\hskip12pt  0  &\hskip12pt  1 \\
\hskip12pt  0  &\hskip2pt -1 &\hskip12pt  0  &\hskip12pt  0  &\hskip2pt -1 &\hskip12pt  0\\
\hskip12pt  0   &\hskip12pt  0  &\hskip12pt  1  &\hskip2pt -1 &\hskip12pt  0  &\hskip12pt  0\\
\hskip2pt -1 &\hskip12pt  0  &\hskip12pt  0  &\hskip12pt  0  &\hskip12pt  0  &\hskip12pt  1\\
\hskip12pt  0  &\hskip2pt -1 &\hskip12pt  0   &\hskip12pt  0  &\hskip12pt  1  &\hskip12pt  0\\
\hskip12pt  0  &\hskip12pt  0  &\hskip12pt  1  &\hskip12pt  1  &\hskip12pt  0  &\hskip12pt  0  
\end{pmatrix}
\end{eqnarray}
\end{subequations}

While it is intuitive to work in bases of L\"owdin orbitals or bonding orbital states, the SALC bases permit the use of the general matrix element theorem and the symmetry analysis based on the point group of the crystal, and to enforce time reversal symmetry on intraband interactions.

\section{Symmetrized exponential functions}

The same procedure can be carried out upon each of the closed sets of EFs for each given shell forming the representation $\Gamma^{EFn}$ of the point group. This generates sets of SEFs which transform as IRs of $D_{6h}$. The equivalence representations of the first two nearest-neighbor shells, and their decompositions, are shown in Table~\ref{tbl:equiv_sef}. The argument presented in the main paper shows the SEFs with $\Gamma_1^-, \Gamma_2^-, \Gamma_6^-, \Gamma_3^+, \Gamma_4^+$, and $\Gamma_5^+$ are forbidden by symmetry.

\begin{table*}
\caption{Decomposition of equivalence representations of Exponential Functions.}
\label{tbl:equiv_sef}
\begin{center}
\begin{tabular}{|c|c|c|c|c|c|c|c|c|c|c|c|c|c|} \hline
$\Gamma(D_{6h})$ & $E$ & $2C_6$ & $2C_3$ & $C_2$ & $3C_2^\prime$ &$3C_2^{\prime\prime}$ & $\imath$ & $2S_3$ & $2S_6$ & $\sigma_h$ & $3\sigma_d$ & $3\sigma_v$ & Decomposition\\  \hline\hline
$\Gamma^{EF1}$ & 6 & 0 & 0 & 0 & 0 & 2 & 0 & 0 & 0 & 6 & 2 & 0 &$\Gamma_1^+\oplus\Gamma_6^+\oplus\Gamma_4^-\oplus\Gamma_5^-$\\ 
$\Gamma^{EF2}$ & 6 & 0 & 0 & 0 & 2 & 0 & 0 & 0 & 0 & 6 & 0 & 2 & $\Gamma_1^+\oplus\Gamma_6^+\oplus\Gamma_3^-\oplus\Gamma_5^-$\\ \hline
\end{tabular}
\end{center}
\end{table*}

There is no clear way to normalize SEFs. For this reason, the invariant material parameters are not determined until we have a systematic way of determining normalized SEFs. In this manuscript, the column vector obtained from projection operators is normalized before constructing the SEFs. The SEFs for graphene up to second-nearest neighbor shells are given below,
\begin{subequations}
\label{eqn:sefs}
\begin{eqnarray}
\mathcal{K}^{\Gamma_1^+}(\bm{k},0)&=&1\\
\mathcal{K}^{\Gamma_1^+}(\bm{k},1)&=&\frac{2}{\sqrt{6}}\left[\cos\left(k_ya\right) + 2\cos\left({1\over2}k_ya\right)\cos\left({\sqrt{3}\over2}k_xa\right)\right]\\
\mathcal{K}^{\Gamma_6^+,1}(\bm{k},1)&=&\frac{2}{\sqrt{3}}\left[\cos\left(k_ya\right) - \cos\left(\frac{1}{2} k_ya\right)\cos\left(\frac{\sqrt{3} }{2} k_xa\right)\right]\\
\mathcal{K}^{\Gamma_6^+,2}(\bm{k},1)&=&-2\sin\left(\frac{1}{2} k_ya\right)\sin\left(\frac{\sqrt{3}}{2} k_xa\right)\\
\mathcal{K}^{\Gamma_4^-}(\bm{k},1)&=&\frac{2i}{\sqrt{6}}\left[\sin\left(k_ya\right) -2\sin\left(\frac{1}{2}k_ya\right)\cos\left(\frac{\sqrt{3}}{2}k_xa\right)\right]\\
\mathcal{K}^{\Gamma_5^-,1}(\bm{k},1)&=&-2i\sin\left(\frac{\sqrt{3}}{2} k_xa\right)\cos\left(\frac{1}{2} k_ya\right) \\
\mathcal{K}^{\Gamma_5^-,2}(\bm{k},1)&=&\frac{-2i}{\sqrt{3}}\left[\sin\left(k_ya\right) + \sin\left(\frac{1}{2}k_ya\right)\cos\left(\frac{\sqrt{3}}{2}k_xa\right)\right] \\
\mathcal{K}^{\Gamma_1^+}(\bm{k},2)&=&\frac{2}{\sqrt{6}}\left[\cos\left(\sqrt{3}\,k_xa\right)+2\cos\left(\frac{\sqrt{3}}{2}k_xa\right)\cos\left(\frac{3}{2}k_ya\right)\right] \\
\mathcal{K}^{\Gamma_6^+,1}(\bm{k},2)&=&-\frac{2}{\sqrt{3}}\left[\cos\left(\sqrt{3}k_xa\right)-\cos\left(\frac{\sqrt{3}}{2}k_xa\right)\cos\left(\frac{3}{2}k_ya\right) \right]\\
\mathcal{K}^{\Gamma_6^+,2}(\bm{k},2)&=&-2\sin\left(\frac{\sqrt{3}}{2}k_xa\right)\sin\left(\frac{3}{2}k_ya\right) \\
\mathcal{K}^{\Gamma_3^-}(\bm{k},2)&=&\frac{2i}{\sqrt{6}}\left[\sin\left(\sqrt{3}\,k_xa\right) -2\sin\left(\frac{\sqrt{3}}{2}k_xa\right)\cos\left(\frac{3}{2}k_ya\right)\right]\\
\mathcal{K}^{\Gamma_5^-,1}(\bm{k},2)&=&-\frac{2i}{\sqrt{3}} \left[\sin\left(\sqrt{3}k_xa\right)+\sin\left(\frac{\sqrt{3}}{2}k_xa\right)\cos\left(\frac{3}{2}k_ya\right) \right] \\
\mathcal{K}^{\Gamma_5^-,2}(\bm{k},2)&=&-i2\sin\left(\frac{3}{2}k_ya\right)\cos\left(\frac{\sqrt{3}}{2}k_xa\right)
\end{eqnarray}
\end{subequations}

\section{Generator matrices}

The last remaining element required in the application of the method of invariants, are the generator matrices. The matrix $\bigl\langle\phi_\nu|H|\phi_\mu\bigr\rangle$ has $d_\mu\times d_\nu$ elements, with element $H_{ij}$ considered as components of the product basis $\bigl|\phi_{\nu,i}\bigr\rangle^\ast\otimes\big|\phi_{\mu,j}\bigr\rangle$. Thus, the matrix, viewed as a column vector, transforms as $D^M(g)=\left(\mathcal{D}^{\nu}(g)^\ast\otimes\mathcal{D}^{\mu}(g)\right)^\dag$. Hence, the generator which transforms as the $k$th component of  IR $\gamma$ can be obtained from the projection operator,
\begin{equation}
\label{eqn:projector2}
\mathcal{P}^\gamma_{kk}=\frac{d_\mu}{|G|}\sum_{g=1}^{|D_{6h}|}\mathcal{D}^\gamma(g)_{kk}^\ast D^M(g)\, .
\end{equation}

The projection operator technique does not determine the phase factor or sign of the generator matrices. The SEFs obtained for graphene are real for the positive parity representations, and purely imaginary for the negative parity representations. We make use of real generator matrices and the following sign convention:
\begin{align}
\label{eqn:generator_phase}
M_{\nu,\mu}^\gamma(n)&=f(\gamma,n){M_{\mu,\nu}^\gamma}(n)^T\, ,\\
\noalign{\vskip3pt}
f(\gamma,n)&=\mbox{sgn}(\gamma)g(n)\, ,\\
\noalign{\vskip3pt}
g(n)&=\left\{\begin{array}{ll}\hskip9pt1&\mbox{if $c(n)$ is real;} \\ -1&\mbox{if $c(n)$ is imaginary.} \end{array}\right.
\end{align}
Given this convention, the requirement of the Hamiltonian to be Hermitian, places the following constraints on the invariant material parameters,
\begin{equation}
\label{eqn:hermitian_constraint}
c_{\mu,\nu}^\gamma(n)=c_{\nu,\mu}^\gamma(n).
\end{equation}
We have left the possibility that the reduced tensor elements may be purely real or imaginary.

The required generators matrices are given below for the construction of the 8-band Hamiltonian involving the $\pi$ and $\sigma$ bands,
{\allowdisplaybreaks\begin{align}
M_{\Gamma_3^+,\Gamma_3^+}^{\Gamma_1^+}&=M_{\Gamma_1^+,\Gamma_1^+}^{\Gamma_1^+}=M_{\Gamma_2^-,\Gamma_2^-}^{\Gamma_1^+}\notag \\
&=M_{\Gamma_4^-,\Gamma_4^-}^{\Gamma_1^+}=M_{\Gamma_2^-,\Gamma_3^+}^{\Gamma_4^-}=M_{\Gamma_1^+,\Gamma_4^-}^{\Gamma_4^-}=1\, , \\
M_{\Gamma_5^-,\Gamma_4^-}^{\Gamma_6^+,1}&=M_{\Gamma_6^+,\Gamma_4^-}^{\Gamma_5^-,1}=\left(\begin{array}{c}0\\ \noalign{\vskip6pt} 1
\end{array}\right)\, ,\\
M_{\Gamma_5^-,\Gamma_4^-}^{\Gamma_6^+,2}&=M_{\Gamma_6^+,\Gamma_4^-}^{\Gamma_5^-,2}=\left(\begin{array}{r}-1\\ \noalign{\vskip6pt} 0
\end{array}\right)\, ,\\
\noalign{\vskip3pt}
M_{\Gamma_1^+,\Gamma_5^-}^{\Gamma_5^-,1}&=M_{\Gamma_1^+,\Gamma_6^+}^{\Gamma_6^+,1}=\begin{pmatrix}1 & 0\end{pmatrix}\, ,\\
M_{\Gamma_1^+,\Gamma_5^-}^{\Gamma_5^-,2}&=M_{\Gamma_1^+,\Gamma_6^+}^{\Gamma_6^+,2}=\begin{pmatrix}0 & 1\end{pmatrix}\, , \\
M_{\Gamma_5^-,\Gamma_5^-}^{\Gamma_6^+,1}&=M_{\Gamma_6^+,\Gamma_5^-}^{\Gamma_5^-,1}=M_{\Gamma_6^+,\Gamma_6^+}^{\Gamma_6^+,1}=\left(\begin{array}{cr}1&\hskip3pt 0\\\noalign{\vskip6pt}0&\hskip3pt -1\end{array}\right)\, ,\\
\noalign{\vskip3pt}
M_{\Gamma_5^-,\Gamma_5^-}^{\Gamma_6^+,2}&=M_{\Gamma_6^+,\Gamma_5^-}^{\Gamma_5^-,2}=M_{\Gamma_6^+,\Gamma_6^+}^{\Gamma_6^+,2}=-\left(\begin{array}{cc}0&\hskip12pt 1\\ \noalign{\vskip6pt} 1&\hskip12pt 0\end{array}\right)\, ,\\
\noalign{\vskip3pt}
M_{\Gamma_6^+,\Gamma_5^-}^{\Gamma_3^-}&=\left(\begin{array}{cc}1&\hskip12pt 0\\ \noalign{\vskip6pt} 0&\hskip12pt 1\end{array}\right)\, ,\\
\noalign{\vskip3pt}
M_{\Gamma_6^+,\Gamma_5^-}^{\Gamma_4^-}&=\left(\begin{array}{cr}0&\hskip3pt-1\\\noalign{\vskip6pt}1&\hskip3pt 0\end{array}\right)\, .
\end{align}}
The double group generator matrices are given below for the construction of the 4-band Hamiltonian involving the $\pi$ band only,
\begin{align}
M_{\Gamma_8^+,\Gamma_8^+}^{\Gamma_1^+}&=M_{\Gamma_7^-,\Gamma_7^-}^{\Gamma_1^+}=M_{\Gamma_7^-,\Gamma_8^+}^{\Gamma_3^-}=
\left(
\begin{array}{cc}
1&\hskip12pt 0\\
\noalign{\vskip6pt}
0&\hskip12pt 1
\end{array}
\right)\, ,\\
M_{\Gamma_7^-,\Gamma_8^+}^{\Gamma_4^-}&=
\left(
\begin{array}{cr}
1&\hskip4pt 0\\
\noalign{\vskip6pt}
0&\hskip4pt -1
\end{array}
\right)\, .
\end{align}

\section{Localized orbital constraints on invariant material parameters}

We now have all the ingredients to construct a Hamiltonian which is invariant under the actions of point group elements and is Hermitian, subject to the constraint described in Eq.~(\ref{eqn:hermitian_constraint}). The general matrix element theorem is applied, together with time-reversal rules, in finding the symmetry-permitted generators. The Bloch sums constructed from the SALC bases serve as basis functions of the Hamiltonian. 

There are some additional constraints which must be imposed on the invariant parameters to ensure localized L\"owdin orbitals on atomic sites. This requires the invariant Hamiltonian, obtained from Eq.~(\ref{eqn:invariant_shell1}), to be equivalent to those obtained from SK formulation under the similarity transformation defined by Eq.~(\ref{eqn:lowdin2salc}), at least for interactions under the two-center approximation.  In the SK formulation, the Hamiltonian may be partitioned into four blocks of $H_{AB}, H_{BA}, H_{AA}$, and $H_{BB}$ if the bases are ordered according to the type of atomic sites. The $H_{AB}, H_{BA}$ blocks describe interactions between localized L\"owdin orbitals on different type of sites, whereas the $H_{AA}, H_{BB}$ blocks describes interactions between localized L\"owdin orbitals on the same type of sites. Specifically, terms involving SEFs of shells coupling the same sites (for example AA, BB) should appear in the appropriate partitions when transformed into the L\"owdin orbital bases. Terms coupling different sites (AB, BA) should appear in the appropriate partitions and have the correct EF dependence required by bond vectors when transformed into the L\"owdin orbital bases. 

Applying the similarity transform defined in Eq.~(\ref{eqn:lowdin2salc}) to the invariant Hamiltonian, the result must only occur in the appropriate quadrant of the Hamiltonian for a given shell. (For example, the nearest neighbor interaction must occur in the $H_{AB}$ and $H_{BA}$ partition after the transformation.) For inter-site interactions, the form of the Hamiltonian must reflect the exponential functions obtained for the appropriate bond vectors. This places further constraints on the material parameters, as detailed in Eq.~(\ref{eqn:local_const}) below. Among the following parameters, those in red are chosen to be independent parameters.

\noindent
For the onsite interaction, we have:
\begin{subequations}
\label{eqn:local_const}
\begin{eqnarray}
c_{\Gamma_2^-,\Gamma_2^-}^{\Gamma_1^+}(0)&=&-{\textcolor{red}c_{\Gamma_3^+,\Gamma_3^+}^{\Gamma_1^+}(0)} \\
c_{\Gamma_1^+,\Gamma_1^+}^{\Gamma_1^+}(0)&=&-{\textcolor{red}c_{\Gamma_4^-,\Gamma_4^-}^{\Gamma_1^+}(0)} \\
c_{\Gamma_6^+,\Gamma_6^+}^{\Gamma_1^+}(0)&=&-{\textcolor{red}c_{\Gamma_5^-,\Gamma_5^-}^{\Gamma_1^+}(0)}.
\end{eqnarray}
For first neighbor interactions, we have:
\begin{eqnarray}
c_{\Gamma_2^-,\Gamma_2^-}^{\Gamma_1^+}(1)&=&-{\textcolor{red}c_{\Gamma_3^+,\Gamma_3^+}^{\Gamma_1^+}(1)} \\
c_{\Gamma_6^+,\Gamma_6^+}^{\Gamma_1^+}(1)&=&-{\textcolor{red}c_{\Gamma_5^-,\Gamma_5^-}^{\Gamma_1^+}(1)} \\
c_{\Gamma_1^+,\Gamma_1^+}^{\Gamma_1^+}(1)&=&-{\textcolor{red}c_{\Gamma_4^-,\Gamma_4^-}^{\Gamma_1^+}(1)} \\
c_{\Gamma_6^+,\Gamma_1^+}^{\Gamma_6^+}(1)&=&-{\textcolor{red}c_{\Gamma_4^-,\Gamma_5^-}^{\Gamma_6^+}(1)} \\
c_{\Gamma_4^-,\Gamma_6^+}^{\Gamma_5^-}(1)&=&-{\textcolor{red}c_{\Gamma_5^-,\Gamma_1^+}^{\Gamma_5^-}(1)} \\
c_{\Gamma_6^+,\Gamma_6^+}^{\Gamma_6^+}(1)&=&{\textcolor{red}c_{\Gamma_5^-,\Gamma_5^-}^{\Gamma_6^+}(1)} \\
c_{\Gamma_5^-,\Gamma_6^+}^{\Gamma_4^-}(1)&=&c_{\Gamma_5^-,\Gamma_5^-}^{\Gamma_1^+}(1)\\ 
c_{\Gamma_3^+,\Gamma_2^-}^{\Gamma_4^-}(1)&=&-c_{\Gamma_3^+,\Gamma_3^+}^{\Gamma_1^+}(1) \\
c_{\Gamma_4^-,\Gamma_1^+}^{\Gamma_4^-}(1)&=&-c_{\Gamma_4^-,\Gamma_4^-}^{\Gamma_1^+}(1) \\
c_{\Gamma_4^-,\Gamma_5^-}^{\Gamma_6^+}(1)&=&c_{\Gamma_5^-,\Gamma_1^+}^{\Gamma_5^-}(1) \\
c_{\Gamma_5^-,\Gamma_6^+}^{\Gamma_5^-}(1)&=&c_{\Gamma_5^-,\Gamma_5^-}^{\Gamma_6^+}(1) 
\end{eqnarray}
For the second-nearest neighbor interaction, we have
\begin{eqnarray}
c_{\Gamma_2^-,\Gamma_2^-}^{\Gamma_1^+}(2)&=&{\textcolor{red}c_{\Gamma_3^+,\Gamma_3^+}^{\Gamma_1^+}(2)} \\
c_{\Gamma_6^+,\Gamma_6^+}^{\Gamma_1^+}(2)&=&{\textcolor{red}c_{\Gamma_5^-,\Gamma_5^-}^{\Gamma_1^+}(2)} \\
c_{\Gamma_1^+,\Gamma_1^+}^{\Gamma_1^+}(2)&=&{\textcolor{red}c_{\Gamma_4^-,\Gamma_4^-}^{\Gamma_1^+}(2)} \\
c_{\Gamma_4^-,\Gamma_6^+}^{\Gamma_5^-}(2)&=&{\textcolor{red}c_{\Gamma_5^-,\Gamma_1^+}^{\Gamma_5^-}(2)} \\
c_{\Gamma_6^+,\Gamma_6^+}^{\Gamma_6^+}(2)&=&-{\textcolor{red}c_{\Gamma_5^-,\Gamma_5^-}^{\Gamma_6^+}(2)} \\
c_{\Gamma_5^-,\Gamma_6^+}^{\Gamma_5^-}(2)&=&0 \\
c_{\Gamma_6^+,\Gamma_1^+}^{\Gamma_6^+}(2)&=&{\textcolor{red}c_{\Gamma_4^-,\Gamma_5^-}^{\Gamma_6^+}(2)} \\
c_{\Gamma_6^+,\Gamma_5^-}^{\Gamma_3^-}(2)&=&{\textcolor{red}c_{\Gamma_5^-,\Gamma_6^+}^{\Gamma_3^-}(2)}.
\end{eqnarray}
\end{subequations}
A Hamiltonian, which is invariant under rotational and time-reversal, can be constructed using the SEFs, generator matrices and invariant material parameters using Eq.\eqref{eqn:invariant_shell1} subject to the constraints. 

Without any spin degrees of freedom, the invariant Hamiltonian for the two-band model is
\begin{align}
\label{eqn:single_ham}
H(\bm{k})&=\left[c_{\Gamma_3^+,\Gamma_3^+}^{\Gamma_1^+}(0){\mathcal{K}^{\Gamma_1^+}(\bm{k},0)}+c_{\Gamma_3^+,\Gamma_3^+}^{\Gamma_1^+}(2){\mathcal{K}^{\Gamma_1^+}(\bm{k},2)}\right]\left(\begin{array}{cc}1&\hskip10pt 0\\ \noalign{\vskip6pt}0&\hskip10pt 1\end{array}\right)\notag\\
\noalign{\vskip3pt}
&\quad+c_{\Gamma_3^+,\Gamma_3^+}^{\Gamma_1^+}(1)\left[{\mathcal{K}^{\Gamma_1^+}(\bm{k},1)}\left(\begin{array}{cr}
1&\hskip2pt 0\\ \noalign{\vskip6pt} 0&\hskip2pt-1\end{array}\right)+{\mathcal{K}^{\Gamma_4^-}(\bm{k},1)}\left(\begin{array}{rc}0&\hskip12pt 1\\ \noalign{\vskip6pt}-1&\hskip12pt 0\end{array}\right)\right]\, .
\end{align}
With the spin degree of freedom, the corresponding Hamiltonian is
\begin{align}
\label{eqn:double_ham}
&H(\bm{k})=\left[c_{\Gamma_8^+,\Gamma_8^+}^{\Gamma_1^+}(0){\mathcal{K}^{\Gamma_1^+}(\bm{k},0)}+c_{\Gamma_8^+,\Gamma_8^+}^{\Gamma_1^+}(2){\mathcal{K}^{\Gamma_1^+}(\bm{k},2)}\right]\left(\begin{array}{cccc}1&\hskip10pt 0&\hskip10pt 0 &\hskip10pt 0\\ \noalign{\vskip3pt}0&\hskip10pt 1&\hskip10pt 0 &\hskip10pt 0\\ \noalign{\vskip3pt}0&\hskip10pt 0&\hskip10pt 1 &\hskip10pt 0\\ \noalign{\vskip3pt} 0&\hskip10pt 0&\hskip10pt 0 &\hskip10pt 1\end{array}\right)\notag \\
\noalign{\vskip3pt}
&+c_{\Gamma_8^+,\Gamma_8^+}^{\Gamma_1^+}(1)\left[{\mathcal{K}^{\Gamma_1^+}(\bm{k},1)}
\left(\begin{array}{ccrr}1&\hskip10pt 0&\hskip2pt 0&\hskip2pt 0\\ \noalign{\vskip3pt} 0&\hskip10pt 1&\hskip2pt 0&\hskip2pt 0\\ \noalign{\vskip3pt} 0&\hskip10pt 0&\hskip2pt -1&\hskip2pt 0\\ \noalign{\vskip3pt}0&\hskip10pt 0&\hskip2pt 0&\hskip2pt -1\end{array}\right)+{\mathcal{K}^{\Gamma_4^-}(\bm{k},1)}\left(\begin{array}{rccr}0&\hskip10pt 0&\hskip10pt 1&\hskip2pt 0\\ \noalign{\vskip3pt} 0&\hskip10pt 0&\hskip10pt 0&\hskip2pt -1\\ \noalign{\vskip3pt} -1&\hskip10pt 0&\hskip10pt 0&\hskip2pt 0\\ \noalign{\vskip3pt} 0&\hskip10pt 1&\hskip10pt 0&\hskip2pt 0\end{array}\right)\right] \notag \\
\noalign{\vskip3pt}
&\quad + c_{\Gamma_8^+,\Gamma_7^-}^{\Gamma_3^-}(2){\mathcal{K}^{\Gamma_3^-}(\bm{k},2)}\left(\begin{array}{cccc}0&\hskip10pt 0&\hskip10pt 1&\hskip10pt 0\\ \noalign{\vskip3pt}0&\hskip10pt 0&\hskip10pt 0&\hskip10pt 1\\ \noalign{\vskip3pt}1&\hskip10pt 0&\hskip10pt 0&\hskip10pt 0\\ \noalign{\vskip3pt}0&\hskip10pt 1&\hskip10pt 0&\hskip10pt 0\end{array}\right)\, ,
\end{align}
with $c_{\Gamma_8^+,\Gamma_7^-}^{\Gamma_3^-}(2)$ imaginary. The approximate relations between the single and double group invariant material parameters are,
\begin{subequations}
\label{eqn:single_double_relation}
\begin{eqnarray}
c_{\Gamma_8^+,\Gamma_8^+}^{\Gamma_1^+}(0)&=&c_{\Gamma_3^+,\Gamma_3^+}^{\Gamma_1^+}(0)\\
c_{\Gamma_8^+,\Gamma_8^+}^{\Gamma_1^+}(1)&=&c_{\Gamma_3^+,\Gamma_3^+}^{\Gamma_1^+}(1)\\
c_{\Gamma_8^+,\Gamma_8^+}^{\Gamma_1^+}(2)&=&c_{\Gamma_3^+,\Gamma_3^+}^{\Gamma_1^+}(2) \\
c_{\Gamma_8^+,\Gamma_8^+}^{\Gamma_3^-}(2)&=&\frac{ic_{\Gamma_8^+(\pi),\Gamma_8^+(\sigma)}^{\Gamma_1^+}(0)\cdot c_{\Gamma_7^-(\pi),\Gamma_7^-(\sigma)}^{\Gamma_1^+}(0)}{(E_{\Gamma_3^+}-E_{\Gamma_6^+})(E_{\Gamma_5^-}-E_{\Gamma_2^-})}c_{\Gamma_8^+(\sigma),\Gamma_7^-(\sigma)}^{\Gamma_3^-}(2) \notag \\
&=&\frac{i\Delta_{so}^2}{(E_{\Gamma_3^+}-E_{\Gamma_6^+})(E_{\Gamma_5^-}-E_{\Gamma_2^-})}c_{\Gamma_5^-,\Gamma_6^+}^{\Gamma_3^-}(2)
\end{eqnarray}
\end{subequations}
where $\Delta_{so}^2=c_{\Gamma_8^+(\pi),\Gamma_8^+(\sigma)}^{\Gamma_1^+}(0)\cdot c_{\Gamma_7^-(\pi),\Gamma_7^-(\sigma)}^{\Gamma_1^+}(0)$. The last relation in Eq.~\eqref{eqn:single_double_relation} is obtained using perturbation theory to treat the mixing between $\sigma$ and $\pi$ by the spin-orbit interaction and the second-nearest neighbor interaction between the $\sigma$ orbitals via the three center interaction between the $\Gamma_5^-$ and $\Gamma_6^+$ states associated with $\mathcal{K}^{\Gamma_3^-}(2)$. The values of these parameters used to produce the dispersion in Fig.~2 of the main paper are,
\begin{eqnarray*}
c_{\Gamma_3^+,\Gamma_3^+}^{\Gamma_1^+}(0)&=&0.2610\:eV\\
c_{\Gamma_3^+,\Gamma_3^+}^{\Gamma_1^+}(1)&=&3.5865\:eV \\
c_{\Gamma_3^+,\Gamma_3^+}^{\Gamma_1^+}(2)&=&0.2131\:eV \\
c_{\Gamma_8^+,\Gamma_7^-}^{\Gamma_3^-}(2)&=&9.622\:\mu eV
\end{eqnarray*}

In comparison to the SK formulation, the second-nearest neighbor interaction contains two more parameters $c_{\Gamma_4^-,\Gamma_5^-}^{\Gamma_6^+}(2)$ and $c_{\Gamma_5^-,\Gamma_6^+}^{\Gamma_3^-}(2)$. These are associated with three-center interactions \cite{stiles97} which are neglected in the SK formulation. The presence of the $c_{\Gamma_5^-,\Gamma_6^+}^{\Gamma_3^-}(2)$ parameter is crucial for explaining the gap at $K/K^\prime$ points due to inter-site spin-orbit interactions, as discussed in the main paper. The hopping parameter for the time-reversal symmetry-breaking term ($\mathcal{K}^{\Gamma_3^-}(\bm{k},2)$) under a periodic magnetic field is due to three-center interactions, and quite different from the two-center-mediated hopping under the SK formulation ($\mathcal{K}^{\Gamma_1^+}(\bm{k},2)$), which breaks the symmetry between electron and hole states. Therefore, any SK formulation of the tight-binding method would not be able to explain the occurrence of the intrinsic gap in graphene.

Assuming $\bm{\kappa}=\bm{k}-\bm{K}_0$ where $\bm{K}_0$ takes on the value at $K$ and $K^\prime$ point, a Taylor expansion of Eq.~\eqref{eqn:double_ham} gives
\begin{equation}
H(\bm{\kappa})=c_1a\left(
\begin{array}{rr}
\mp\kappa_x\sigma_0 & i\kappa_y\sigma_3\\
\noalign{\vskip6pt}
-i\kappa_y\sigma_3 & \pm\kappa_x\sigma_0
\end{array}
\right)\pm c_2\left(
\begin{array}{cc}
0 &\hskip6pt \sigma_0\\
\noalign{\vskip6pt}
\sigma_0 &\hskip6pt 0
\end{array}
\right)\, .
\end{equation}
where the choice of `+' and `-' correspond to expansion at $K$ and $K^\prime$ respectively. After a transformation to atomic site basis, we obtain Eq.~(6) of the main paper.

It should be recognized that the requirement of a localized L\"owdin bases may be broken by the inter-site spin orbit interaction. In the context of the SK formulation, the key to incorporating the spin-orbit interaction is to understand the role of intra-site ($AA$,$BB$) and inter-site ($AB$,$BA$) spin-orbit interactions. The intra-site spin-orbit interaction modifies the zone center energies, leading to spin splitting in single group states with $\Gamma_5^-$ and $\Gamma_6^+$ symmetry. These correspond to the symmetry-allowed diagonal elements, and there is no consequent modification of the $\bm{k}$-dependence. In contrast, the inter-site spin-orbit interaction appears in the $AB/BA$ partitions of the Hamiltonian, though this does not incur a $\bm{k}$-dependence. In other words, it has an intra-site $\bm{k}$-dependence. Constraints based on the requirement of localized atomic wave functions thus may be broken by the inter-site spin-orbit interaction. This is, of course, subject to the invariant requirement under the action of the point group. In the case of graphene, terms of this nature are not symmetry invariant.

\section{Berry phase under cell-periodic magnetic fields}
The work of Haldane needs to be reinterpreted in light of the symmetry analysis. First of all, Haldane is correct to assert that the symmetry group of the graphene crystal is not affected by the introduction of a cell-periodic magnetic field. The cell-periodic field is external and not part of the crystal. It remains fixed to the coordinate system. Any symmetry operation of the graphene space group would leave the crystal invariant under the space with embedded cell-periodic magnetic field.  However, any closed hopping circuit (part of the crystal), and the associated Berry phase would transform under the action of the point group. If they encompass a complete primitive cell, then it should be invariant and the Berry phase should be zero because of the cell-periodic nature of the external field. If they enclose areas covering only half of a primitive cell, then the associated Berry phase would change under the action of the point group of the crystal.
\begin{figure}[htbp]
\begin{center}
\includegraphics[width=8.5cm]{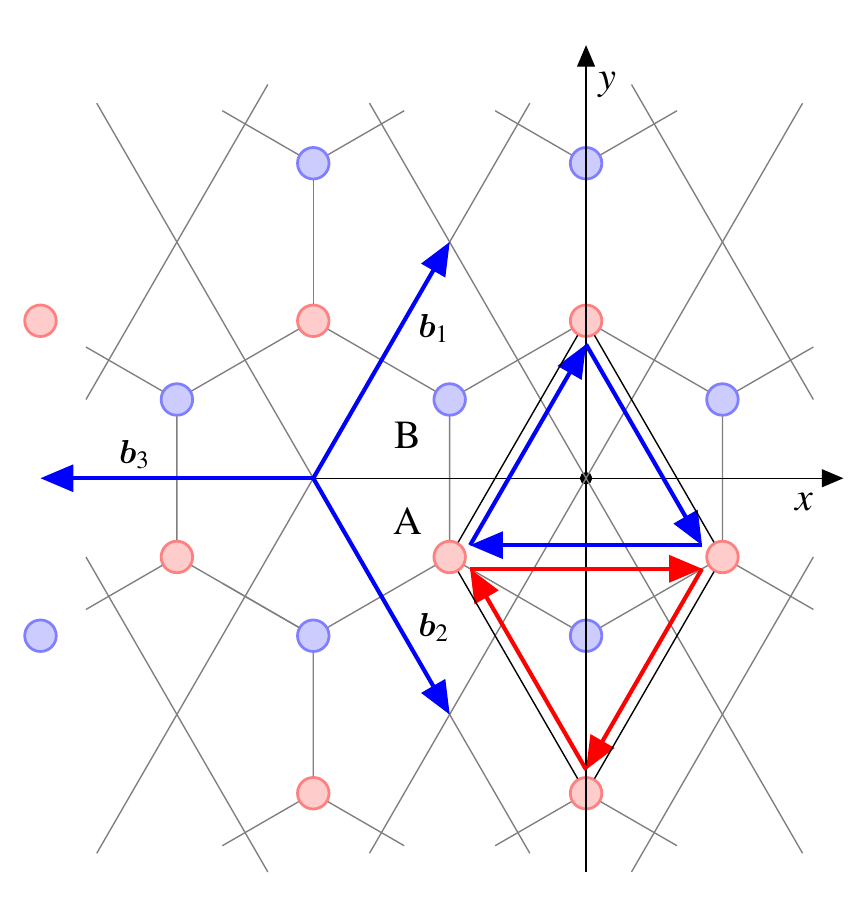}
\caption{Two closed hopping path for second nearest neighbor interaction. The opposing field in the two half of the primitive cell give rise to a Berry phase of $\exp(+\mathrm{i}\phi)$ and $\exp(-\mathrm{i}\phi)$ respectively for the two closed paths. }
\label{fig3}
\end{center}
\end{figure}

Since the presence of a cell-periodic magnetic field does not change the space group, the angular dependent part of interaction matrices (generators in the context of method of invariant) are the same. The Berry phase is added to the relevant terms in the SEF's given a particular closed hopping path. This does not affect the onsite and the nearest-neighbor interaction, as the hopping path encloses the entire primitive cell. There is no extra phase factor picked up in such closed hopping path. In the case of second-nearest neighbor interaction, the time reversal symmetry is broken. For both the $\mathcal{K}^{\Gamma_1^+}(\bm{k},2)$ and $\mathcal{K}^{\Gamma_3^-}(\bm{k},2)$, the hopping path consists of two closed circuits as illustrated in Fig.~3.  To incorporate the Berry phase associated with the cell-periodic magnetic field, we break each of the SEFs into its constituent in terms of the closed paths. We define the lattice vectors $b_1, b_2, b_3$ as indicated in Fig.~3. Then the two set of closed circuits correspond to hopping described by $b_1, b_2, b_3$ and $-b_3, -b_1, -b_2$. We can express the SEFs as
\begin{subequations}
\begin{eqnarray}
\mathcal{K}^{\Gamma_1^+}(\bm{k},2)&=&\sum_{i=1}^3\exp(i\bm{k}\cdot \bm{b}_i)+\sum_{i=1}^3\exp(-i\bm{k}\cdot \bm{b}_i)\notag=A+B\\
\mathcal{K}^{\Gamma_3^-}(\bm{k},2)&=&\sum_{i=1}^3\exp(i\bm{k}\cdot \bm{b}_i)-\sum_{i=1}^3\exp(-i\bm{k}\cdot \bm{b}_i)\notag =A-B
\end{eqnarray}
\end{subequations}
The magnetic fluxes through the two circuits are equal in magnitude but opposite in sign in order to have net zero flux over the primitive cell. Thus, the Aharonov-Bohm phase for each of the circuits may be written as $e^{\pm i\phi}$. Taking into account the phase factors, the SEFs may be written as
\begin{subequations}
\label{eqn:berry_phase}
\begin{eqnarray}
\mathcal{K}^{\Gamma_1^+}&(\bm{k},2)\Rightarrow Ae^{i\phi}+Be^{-i\phi}=\cos(\phi)(A+B)+i\sin(\phi)(A-B) \notag \\
&=\cos(\phi)\mathcal{K}^{\Gamma_1^+}(\bm{k},2)+i\sin(\phi)\mathcal{K}^{\Gamma_3^-}(\bm{k},2)\\
\mathcal{K}^{\Gamma_3^-}&(\bm{k},2)\Rightarrow Ae^{i\phi}-Be^{-i\phi}=\cos(\phi)(A-B)+i\sin(\phi)(A+B) \notag \\
&=\cos(\phi)\mathcal{K}^{\Gamma_3^-}(\bm{k},2)+i\sin(\phi)\mathcal{K}^{\Gamma_1^+}(\bm{k},2) \label{eqn:berry_phaseb}
\end{eqnarray}
\end{subequations}
where the expressions shown in the two equations have overall transformation properties of $\Gamma_1^+$ and $\Gamma_3^-$, respectively, as before ($\cos\phi$ and $\sin\phi$ transform according to $\Gamma_1^+$ and $\Gamma_3^-$, respectively). They should be paired with their respective generators in constructing the Hamiltonian using the method of invariants. This can be easily done in either the single or double group case by replacing $\mathcal{K}^{\Gamma_1^+}(\bm{k},2)$ (single group) in Eq.~\eqref{eqn:single_ham} or $\mathcal{K}^{\Gamma_1^+}(\bm{k},2)$ and $\mathcal{K}^{\Gamma_3^-}(\bm{k},2)$ (double group) in Eq.~\eqref{eqn:double_ham} using the expressions in Eq.~(\ref{eqn:berry_phase}). One can see that these expression reduce to the normal SEFs in the absence of the cell-periodic magnetic field ($\phi=0$).

Under the single group, the contribution from Eq.~(\ref{eqn:berry_phaseb}) is absent because of the forbidden generator $M_{\Gamma_3^+,\Gamma_2^-}^{\Gamma_3^-}$. The third term in Eq.~(1) of Haldane is not invariant under symmetry because $\sin\phi\sum_i \sin(k\cdot b_i)$ transforms as $\Gamma_1^+$ and the generator of $\sigma_3$ is not appropriate. This term shares the same generator as the first term, and the combination gives a corrected Haldane Hamiltonian for single group
\begin{eqnarray}
H(\bm{k})&=&2t_2\sum_{i=1}^3\left[\cos\phi\cos(k\cdot b_i)-\sin\phi\sin(k\cdot b_i)\right]\sigma_0+t_1\sum_{i=1}^3\left[\cos(k\cdot a_i)\sigma_1+\sin(k\cdot a_i)\sigma_2\right],\qquad
\end{eqnarray}
where the two terms correspond to second-nearest and nearest-neighbor hopping. The system remains gapless at $K/K^\prime$ without electron spin (other than those of Landau level separation).
 
The double group Hamiltonian can be transformed to localized orbital basis and allow a extension of the Haldane model by including spin. This yield:
\begin{eqnarray}
H(\bm{k}&)&=2t_2\sum_{i=1}^3\left[\cos\phi\cos(k\cdot b_i)-\sin\phi\sin(k\cdot b_i)\right]\sigma_0\otimes\sigma_0 \notag\\
&&+2t_2^\prime\sum_{i=1}^3\left[\sin\phi\cos(k\cdot b_i)+\cos\phi\sin(k\cdot b_i)\right]\sigma_{3}\otimes\sigma_{3} \notag \\
&&+t_1\sum_{i=1}^3\left[\cos(k\cdot a_i)\sigma_1\otimes\sigma_0+\sin(k\cdot a_i)\sigma_2\otimes\sigma_0\right]
\end{eqnarray}
The condition $\phi=0$ yields the tight-binding Hamiltonian without a field. It should be emphasized that the second-nearest neighbor hopping parameter responsible for the removal of electron/hole symmetry ($t_2/c_{\Gamma_3^+,\Gamma_3^+}^{\Gamma_1^+}(2))$ is quite different from the hopping parameter responsible for intrinsic gap ($t_2^\prime$).
